\documentclass[preprint,authoryear,12pt]{elsarticle}

\usepackage{amssymb,amsmath}
\usepackage{rotating}
\journal{Icarus}

\begin{document}

\begin{frontmatter}

  \title{Near Earth Asteroids with measurable Yarkovsky effect}

  \author[jpl,spacedys]{D. Farnocchia}
  \ead{Davide.Farnocchia@jpl.nasa.gov}
  \author[jpl]{S. R. Chesley}
  \author[prague]{D. Vokrouhlick\'{y}}
  \author[unipi]{A. Milani}
  \author[unipi]{F. Spoto}

   \address[jpl]{Jet Propulsion Laboratory/Caltech, 4800 Oak Grove
            Drive, Pasadena, CA 91109, USA}
   \address[spacedys]{SpaceDyS, Via Mario Giuntini 63, 56023 Cascina,
    Pisa, Italy}
  \address[prague]{ Institute of Astronomy, Charles University, V Hole{\v
              s}ovi{\v c}k{\'a}ch 2, CZ-18000 Prague 8, Czech Republic}
  \address[unipi]{Department of Mathematics, University of Pisa, Largo
    Pontecorvo 5, 56127 Pisa, Italy}

\begin{abstract}
  We seek evidence of the Yarkovsky effect among Near Earth Asteroids
  (NEAs) by measuring the Yarkovsky-related orbital
  drift from the orbital fit.
  To prevent the occurrence of unreliable detections we employ a
  high precision dynamical model, including the Newtonian
  attraction of 16 massive asteroids and the planetary relativistic
  terms, and a suitable astrometric data treatment.
  We find 21 NEAs whose orbital fits show a measurable orbital
  drift with a signal to noise ratio (SNR) greater than 3. 
  The best determination is for asteroid (101955) 1999 RQ$_{36}$,
  resulting in the recovery of one radar apparition and an orbit
  improvement by two orders of magnitude.
  In addition, we find 16 cases with a lower SNR that, despite
  being less reliable, are good candidates for becoming stronger
  detections in the future.
  In some cases it is possible to constrain physical
  quantities otherwise unknown by means of the detected orbital
  drift.
  Furthermore, the distribution of
  the detected orbital drifts shows an excess
  of retrograde rotators that can be connected to the delivery
  mechanism from the most important NEA feeding resonances and allows
  us to infer the distribution for NEAs obliquity.
  We discuss the implications of the Yarkovsky effect for impact
  predictions. In particular, for asteroid (29075) 1950 DA our results
  favor a retrograde rotation that would rule out an impact in 2880.

\end{abstract}


\end{frontmatter}

\section{Introduction}
It is well known that nongravitational forces should be considered as
important as collisions and gravitational perturbations for the
overall understanding of asteroid evolution \citep{bottke}. The most
important nongravitational perturbation is the Yarkovsky effect, which
is due to radiative recoil of anisotropic thermal emission and causes
asteroids to undergo a secular semimajor axis drift $da/dt$. Typical
values of $da/dt$ for sub-kilometer NEAs are $10^{-4}$--$10^{-3}$
au/Myr \citep{vok_2000}.

The Yarkovsky acceleration depends on several physical quantities such
as spin state, size, mass, shape, and thermal properties
\citep{vok99}. Furthermore, \citet{rozitis} show that surface
roughness also plays an important role by enhancing the Yarkovsky
related semimajor axis drift by as much as tens of per cent. Though no
complete physical characterization is typically available to compute
the Yarkovsky acceleration based on a thermophysical model, the
orbital drift may be detectable from an astrometric
dataset. As a matter of fact, a purely gravitational dynamics could
result in an unsatisfactory fit to the observational data. This is
especially true when extremely accurate observations are available,
e.g., radar observations, or when the observational dataset spans a
long time interval thus allowing the orbital drift to accumulate and
become detectable.

Until recently, the Yarkovsky effect has been measured directly only
in three cases, (6489) Golevka \citep{golevka}, (152563) 1992 BF
\citep{1992BF}, and recently for (101955) 1999 RQ$_{36}$
\citep{RQ36}. For both Golevka and 1999 RQ$_{36}$ the Yarkovsky
perturbation must be included to fit accurate radar observations
spanning three apparitions. For 1992 BF the Yarkovsky effect is needed
to link 4 precovery observations of 1953. Furthermore, in the case of
1999 RQ$_{36}$ the available physical characterization, along with the
estimate of the Yarkovsky effect, allows the estimate of the asteroid's
bulk density.

\citet{nugent} find 54 detections of semimajor axis drift by
performing a search for semimajor axis drift among NEAs similar to the
one presented in this paper. However, there are differences in the
observational data treatment, in the modeling, and in the selection
filters. A description of the differences and a comparison of the
results is contained in Sec.~\ref{s:comparison}. \citet{nugent_wise}
use WISE-derived geometric albedos and diameters to predict orbital
drifts for 540 NEAs. Even if none of these objects has an
observational record that allows one to measure the predicted orbital
drift, the authors list upcoming observing opportunities that may
reveal the Yarkovsky signal.

The Yarkovsky effect plays an important role for orbital predictions
such as those concerning Earth impacts. In particular, when an
asteroid has an exceptionally well constrained orbit, the Yarkovsky
effect may become the principal source of
uncertainty. \citet{milani_rq36} show how the size of the semimajor
axis drift along with its uncertainty modifies impact predictions for
the next century for 1999 RQ$_{36}$. The cumulative impact probability
is approximately $10^{-3}$, while a Yarkovsky-free propagation would
rule out any impact event. \citet{RQ36} improve the $da/dt$ estimate
by means of September 2011 Arecibo radar measurements and find a
cumulative impact probability approximately $4\times 10^{-4}$. Another
remarkable case is (99942) Apophis. Though only a marginal $da/dt$
estimate is available, \citet{giorgini08} and \citet{apophis} prove
that the occurrence of an impact in 2036 is decisively driven by the
magnitude of the Yarkovsky effect. On the longer term,
\citet{giorgini02} show that an impact between asteroid (29075) 1950
DA and the Earth in 2880 depends on the accelerations arising from
thermal re-radiation of solar energy absorbed by the asteroid.

\section{Methodology}
\subsection{Yarkovsky modeling and determination}\label{s:yarko}
The Yarkovsky effect depends on typically unknown physical
quantities. As the primary manifestation is a semimajor axis drift, we
seek a formulation depending on a single parameter to be determined as
a result of the orbital fit.  To bypass the need of physical
characterization we used a comet-like model \citep{marsden} for
transverse acceleration $a_t=A_2 g(r)$, where $g$ is a suitable
function of the heliocentric distance $r$ and $A_2$ is an unknown
parameter. To determine $A_2$ we used a 7-dimensional differential
corrector: starting from the observational dataset we simultaneously
determine a best-fitting solution for both the orbital elements and
$A_2$.

Once $A_2$ is determined from the orbital fit we estimate
semimajor axis drift by means of Gauss' perturbative equations:
\begin{equation}
\dot a=\frac{2a\sqrt{1-e^2}}{nr}A_2 g(r)\ 
\end{equation}
where $a$ is the semimajor axis, $e$ is the eccentricity and $n$ is
the mean motion. By averaging we obtain
\begin{equation}
\bar{\dot a}=
\frac{a\sqrt{1-e^2}A_2}{\pi}\int_0^T \frac{g(r)}{r}
dt=\frac{A_2}{\pi na}\int_0^{2\pi}r g(r) df
\end{equation}
where $T$ is the orbital period and $f$ is the true anomaly. 
Let us now assume $g(r)=(r_0/r)^d$, where $r_0$ is a normalizing
parameter, e.g., we use $r_0=1$ au.  In this case the semimajor axis
drift is
\begin{equation}
\bar{\dot a}=\frac{A_2 (1-e^2)}{\pi
  n}\left(\frac{r_0}{p}\right)^d\int_0^{2\pi} (1+e\cos f)^{d-1} df\ .
\end{equation}
By Taylor expansion, we have
\begin{equation}
  \int_0^{2\pi}(1+e\cos f)^{d-1} df= \sum_{k=0}^\infty \binom{d-1}{k}e^k 
  \int_0^{2\pi}\cos^k f df\ .
\end{equation}
The odd powers of the cosine average out, so we obtain
\begin{equation}
\bar{\dot a}=\frac{2 A_2 (1-e^2)}{n}\left(\frac{r_0}{p}\right)^d J(e,d)
\end{equation}
where
\begin{equation}
J(e,d)=\sum_{k=0}^{\infty}\alpha_k e^{2k}\ \ ,\ \
\alpha_k=\binom{d-1}{2k}\binom{2k}{k}\frac{1}{2^{2k}}\ .
\end{equation}
The ratio
\begin{equation}\label{eq:recurr}
\frac{\alpha_{k+1}}{\alpha_k}=\left(1-\frac{d+1}{2k+2}\right)
\left(1-\frac{d}{2k+2}\right)
\end{equation}
is smaller than 1 for $d>0$ and $k$ large enough. Therefore,
$\alpha_k$ are bounded and $J(e,d)$ is convergent for any
$e<1$. Equation~\eqref{eq:recurr} can be used to recursively compute
$\alpha_k$ starting from $\alpha_0 = 1$. For integer $d$ the series
$J$ is a finite sum that can be computed analytically, e.g., $J(e,2) =
1$ and $J(e,3) = 1+0.5e^2$.


The proper value of $d$ is not easily determined. From \citet{vok98},
we have
\begin{equation}\label{eq:diurn}
a_t\simeq \frac{4(1-A)}{9}\Phi(r)
f(\Theta) \cos\gamma\ \ ,\ \ f(\Theta)=\frac{0.5\Theta}{1+\Theta+0.5\Theta^2}
\end{equation}
for the Yarkovsky diurnal component (which is typically dominant) ,
where $A$ is the Bond albedo, $\Theta$ is the thermal parameter,
$\gamma$ is the obliquity, and $\Phi(r)$ is the standard radiation
force factor, which is inversely proportional to the bulk density
$\rho$, the diameter $D$, and $r^2$. The thermal parameter $\Theta$ is
related to the thermal inertia $\Gamma$ by means of the following
equation
\begin{equation}\label{eq:thermal}
  \Theta = \frac{\Gamma}{\varepsilon\sigma
    T_*^3}\sqrt{\frac{2\pi}{P}}
\end{equation}
where $\varepsilon$ is the emissivity, $\sigma$ is the Boltzmann's
constant, $T_*$ is the subsolar temperature, and $P$ is the rotation
period.  In this paper we use $d=2$ to match the level of absorbed
solar radiation. Then, from Eq.~\eqref{eq:diurn} we have that
\begin{equation}\label{eq:A2}
  A_2 \simeq \frac{4(1-A)}{9}\Phi(1 \text{au})
  f(\Theta) \cos\gamma\ .
\end{equation}
However, as $T_* \propto r^{-0.5}$ we have that
$\Theta \propto r^{1.5}$, therefore the best value of $d$ depends on
the object's thermal properties:
\begin{itemize}
\item for $\Theta \gg 1$ we obtain $f \propto r^{-1.5}$, which gives $d =
  3.5$;
\item for $\Theta \ll 1$ we obtain $f \propto r^{1.5}$, which gives $d
  = 0.5$.
\end{itemize}
These are limit cases, the true $d$ is always going to be between
them. As a matter of fact, it turns out that most NEAs, whose rotation
period is not excessively large and whose surface thermal inertia is
not excessively small or large, have typically values of $\Theta$ near
unity or only slightly larger, and we can thus expect $d$ value in the
range 2--3. As an example, \citet{RQ36} show that for 1999 RQ$_{36}$
the best match to the Yarkovsky perturbation computed by using a
linear heat diffusion model is $d=2.75$.

What matters to us is that $da/dt$ does not critically depend on the
chosen value of $d$. As an example for asteroid 1999 RQ$_{36}$ we have
that $da/dt=(-18.99 \pm 0.10)\times 10^{-4}$ au/Myr for $d=2$ and
$da/dt=(-19.02 \pm 0.10)\times 10^{-4}$ au/Myr for $d=3$. Another
example is Golevka, for which we obtain $da/dt=(-6.62 \pm 0.64)\times
10^{-4}$ au/Myr for $d=2$ and $da/dt=(-6.87 \pm 0.66)\times 10^{-4}$
au/Myr for $d=3$. In both cases the difference in $da/dt$ due to the
different values assumed for $d$ is well within one standard
deviation.

\subsection{Dynamical model}
To consistently detect the Yarkovsky effect we need to account for the
other accelerations down to the same order of magnitude. For a
sub-kilometer NEA, typical values of $a_t$ range from $10^{-15}$ to
$10^{-13} \mbox{au}/\mbox{d}^2$.

Our N-body model includes the Newtonian accelerations of the Sun,
eight planets, the Moon, and Pluto that are based on JPL's DE405
planetary ephemerides \citep{de405}.  Furthermore, we added the
contribution of 16 massive asteroids, as listed in
Table~\ref{table:masses}.

\begin{table}
  \caption{Gravitational parameters of perturbing
    asteroids. The masses
    of Ceres, Pallas, and Vesta are from \citet{standish}, the ones of
    Euphrosyne and Herculina are from the Institute of Applied Astronomy of RAS, St. Petersburg, Russia
    (http://www.ipa.nw.ru/PAGE/DEPFUND/LSBSS/engmasses.htm), the mass of
    Juno is from \citet{konopliv}, all the others are from \citet{baer_masses}.}
\label{table:masses}
\centering
\begin{tabular}{lc}
\hline\hline
Asteroid & Gm [km$^3$/s$^2$]\\
\hline
(1) Ceres      & 63.200\\
(2) Pallas     & 14.300\\
(4) Vesta      & 17.800\\
(10) Hygea      & 6.0250\\
(29) Amphitrite & 1.3271\\ 
(511) Davida     & 3.9548\\
(65) Cybele     & 1.0086\\
(9) Metis      & 1.3669\\
(15) Eunomia    & 2.2295\\
(31) Euphrosyne & 1.1280\\
(52) Europa     & 1.2952\\
(704) Interamnia & 4.7510\\
(16) Psyche     & 1.7120\\
(3) Juno       & 1.9774\\
(532) Herculina  & 1.5262\\
(87) Sylvia     & 1.3138\\
\hline
\end{tabular}
\end{table}

\begin{figure*}
  \centering
  \includegraphics[width=8cm]{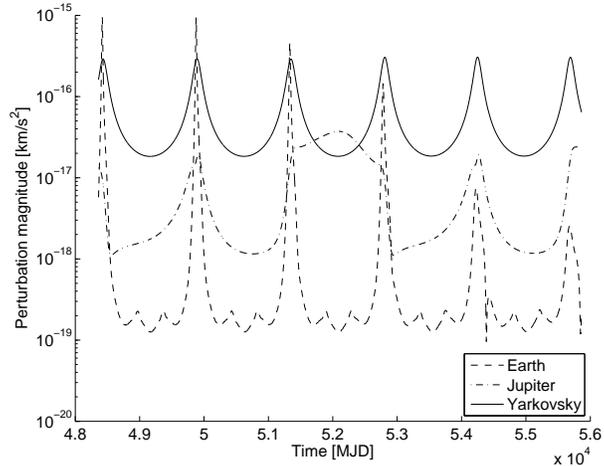}
  \caption{Relativistic accelerations of the Earth and Jupiter as they formally
appear in the EIH equations of motion compared to
    the transverse Yarkovsky acceleration acting on Golevka.}
  \label{fig:plarel}
\end{figure*}

We used a full relativistic force model including the contribution of
the Sun, the planets, and the Moon. Namely, we used the
Einstein-Infeld-Hoffman (EIH) approximation as described in
\citet{moyer} or \citet{will}. As already noted in \citet{RQ36}, the
relativistic term of the Earth should not be neglected because of
significant short range effects during Earth approaches that NEAs may
experience. For asteroids with a large perihelion distance such as
Golevka also Jupiter's term could be relevant. Figure~\ref{fig:plarel}
compares the relativistic accelerations of the Earth and Jupiter as
they formally appear in the EIH equations of motion to the transverse
Yarkovsky acceleration acting on Golevka.

\subsection{Observational error model}\label{s:cbm}
The successful detection of the Yarkovsky effect as a result of the
orbital fit strongly depends on the quality of the observations
involved. In particular, the availability of radar data is often
decisive due to the superior relative accuracy of radar data with
respect to optical ones. Moreover, radar measurements are orthogonal
to optical observations: range and range rate vs. angular position in
the sky.

Since the Yarkovsky effect acts as a secular drift on semimajor axis
we have a quadratic effect in mean anomaly: the longer the time span
the stronger the signal. However, the presence of biases in historical
data and unrealistic weighting of observations may lead to inaccurate
results. To deal with this problem we applied the debiasing and
weighting scheme described in \citet{cbm10}.  This scheme is a valid
error model for CCD observations, while for pre-CCD data the lack of
star catalog information and the very uneven quality of the
observations represents a critical problem. In these cases the
occurrence of unrealistic nominal values for Yarkovsky model
parameters presumably point to bad astrometric treatments and have to
be rejected.

To prevent outliers from spoiling orbital fits, we applied the outlier
rejection procedure as described in \citet{carpino}.

Besides the astrometric treatment described above, in the following
cases we applied an ad hoc observation weighting:
\begin{itemize}
\item 1999 RQ$_{36}$: as already explained by \citet{RQ36}, in some
  cases there are batches containing an excess of observations from a
  single observatory in a single night. To reduce the effect of these
  batches to a preferred contribution of 5 observations per night, we
  relaxed the weight by a factor $\sqrt{N/5}$, where $N$ is the number
  of observations contained in the batch.
\item 1992 BF: as the precovery observations of 1953 have been
  carefully remeasured in \cite{1992BF}, these observations were given
  a weight 0.5'' in right ascension and 1'' in declination.
\item Apollo: the large dataset available for Apollo contains
  observations going back to 1930. Many pre-CCD era observation
  batches show unusually high residuals, especially during close Earth
  approches. To lower the effect of non-CCD observation, we used
  weights 10'' for observations from 1930 to 1950 and 3'' from 1950 to
  1990.
\item 1989 ML: the discovery apparition contains observations from
  Palomar Mountain showing large residuals whether or not the
  Yarkovsky perturbation is included in the model. Even if this
  apparition increases the observed arc by three years only, we felt
  it safer to weight the corresponding observations 3''.
\end{itemize}

\subsection{Treatment of precovery observations}\label{s:precov}
\begin{figure*}[ht!]
  \centering
  \includegraphics[width=6.8cm]{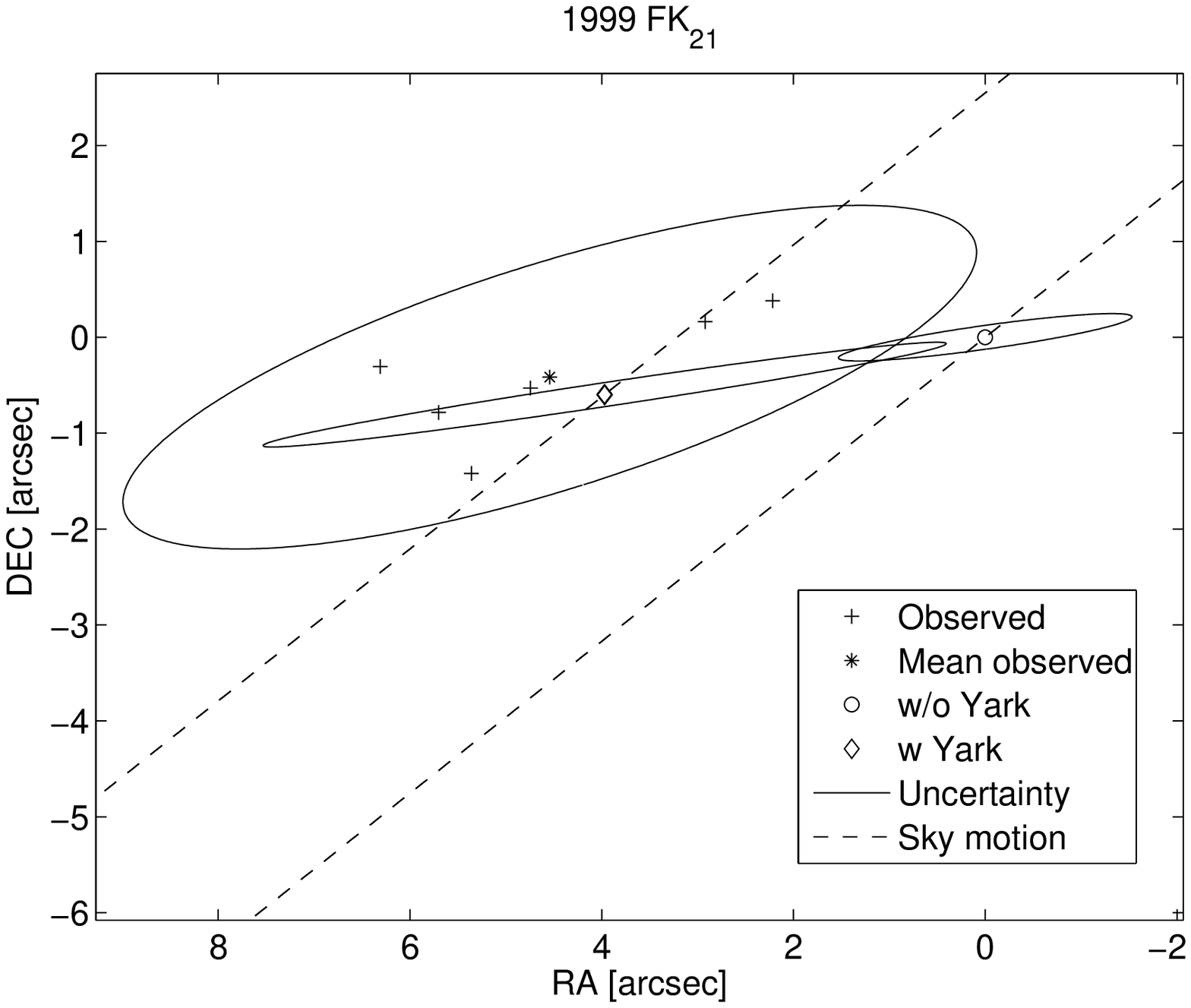}\includegraphics[width=6.8cm]{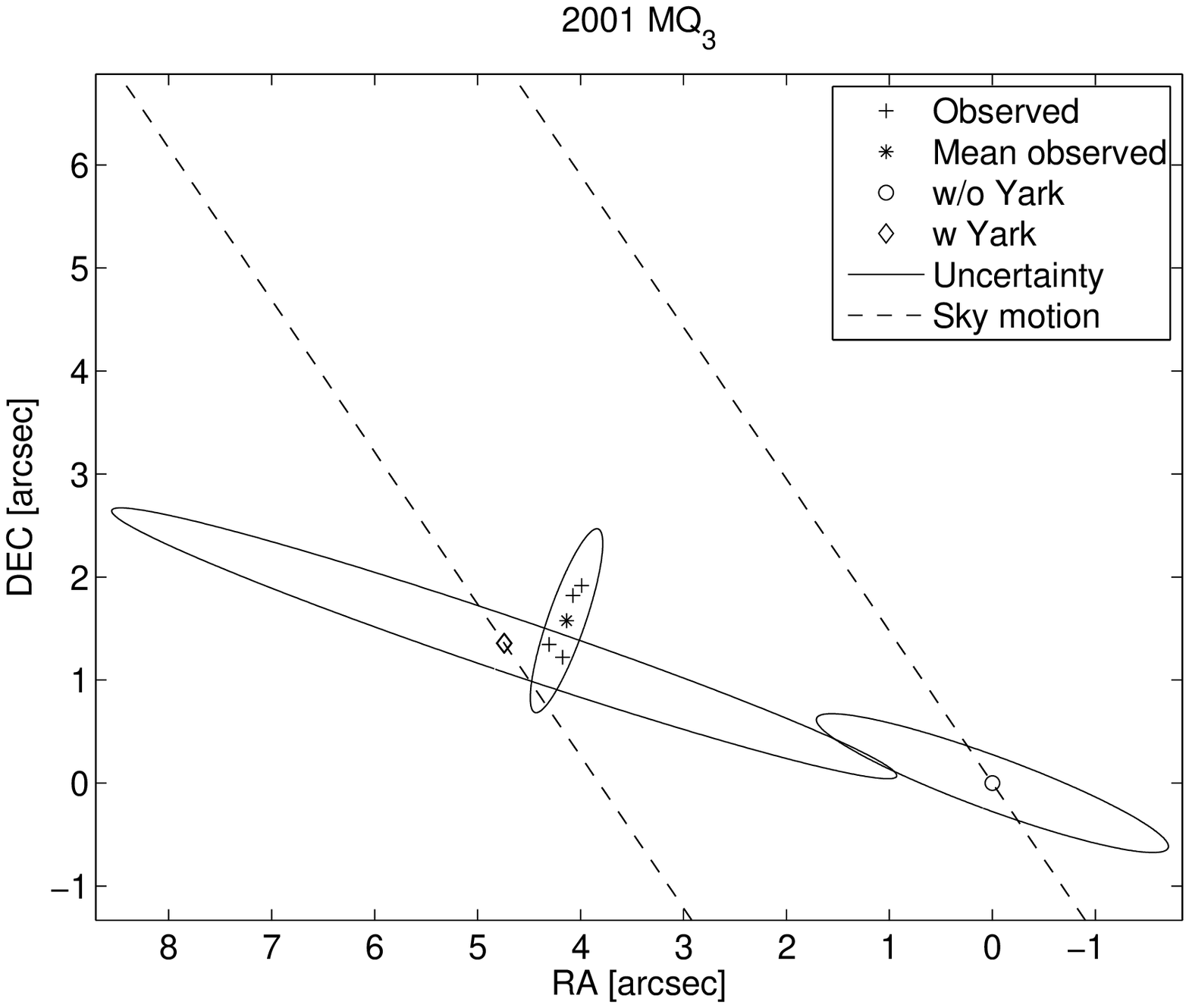}
  \includegraphics[width=6.8cm]{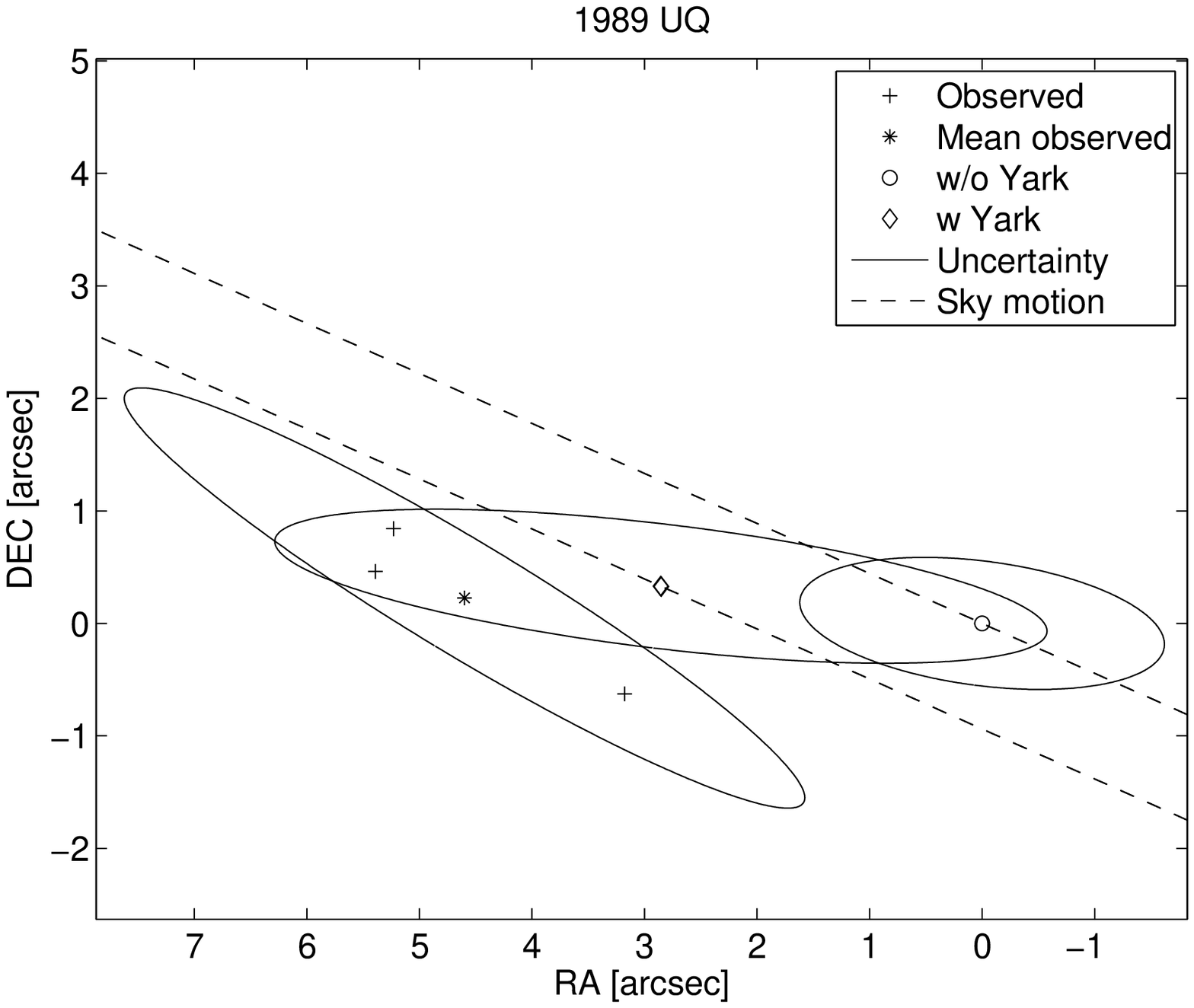}\includegraphics[width=6.8cm]{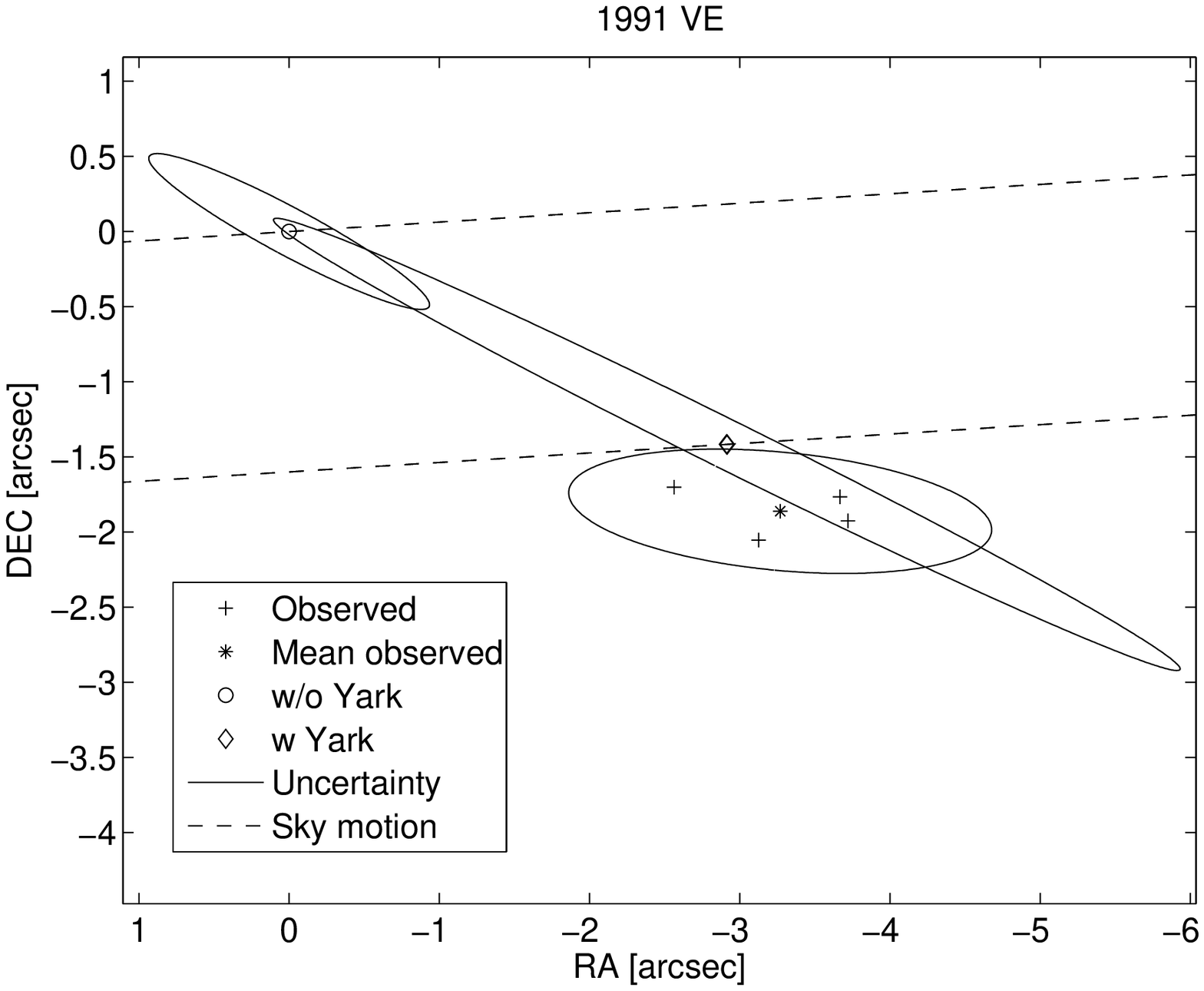}
  \caption{Observed position, average observed position, and predicted
    postfit position with the Yarkovsky perturbation in the dynamics
    with respect to the predictions without the Yarkovsky perturbation
    in the dynamics.  The uncertainty ellipses for the mean observed
    position and the predicted positions correspond to the 3-$\sigma$
    level. The dashed lines represent the predicted motion in the sky
    of the asteroid in the precovery apparition. Circles are the
    predicted positions, crosses are the measured positions, stars are
    the mean of the measured positions, while diamonds are the
    predicted positions when the Yarkovsky perturbation is included in
    the dynamics. The origin is arbitrarily set to the non-Yarkovsky
    prediction.}
  \label{fig:precov}
\end{figure*} 
There are a few cases where the Yarkovsky signal is mainly contained
in few isolated precovery observations. This is the case of the
already mentioned asteroid 1992 BF, which has 4 isolated observations
in 1953 from Palomar Mountain DSS. Other cases are
\begin{itemize}
\item 1999 FK$_{21}$, which has 6 isolated observations in 1971 from
  Palomar Mountain;
\item 2001 MQ$_3$, which has 4 isolated observations in 1951 from
  Palomar Mountain DSS;
\item 1989 UQ, which has 3 isolated observations in 1954 from
  Palomar Mountain;
\item 1991 VE, which has 4 isolated observations in 1954 from
  Palomar Mountain DSS.
\end{itemize}
For all these cases it would be desirable to remeasure the precovery
observations as done for 1992 BF in \citet{1992BF}, where precovery
observations were corrected by an amount up to 3.1''. For this reason
we conservatively gave weights 3'' to the precovery observations of
the four asteroids above.

Besides the conservative weighting, we ruled out clock error as a
possible cause of the Yarkovsky signal. Figure~\ref{fig:precov} shows
the scenario for the four mentioned asteroids during the precovery
apparition.  We can see that it is not
possible to match the observations by translating the non-Yarkovsky
uncertainty ellipse on the along track direction. The Yarkovsky
solution produces a shift in the weak direction that give a better
match to the observations, in particular when we take the average of
the observed positions.

\subsection{Filtering spurious results}\label{s:filter}
To assess the reliability of the Yarkovsky detections we computed an
expected value for $A_2$ starting from the 1999 RQ$_{36}$ case, which
is the strongest and most reliable detection, by scaling according to
\eqref{eq:A2}
\begin{equation}\label{eq:scaling}
\left(A_2\right)_{exp} = \left(A_2\right)_{RQ36}
\frac{D_{RQ36}}{D}\ .
\end{equation}
For diameter $D$ we used either the known value when available or an
assumed value computed from the absolute magnitude $H$ according to
the following relationship \citep{pravec}:
\begin{equation}\label{eq:diameter}
D=1329 \text{ km } \times \frac{10^{-0.2H}}{\sqrt{p_V}}
\end{equation}
where $p_V$ is the albedo, assumed to be 0.154 if unknown, in
agreement with the Palermo scale computation \citep{palermo}.

Some physical properties of 1999 RQ$_{36}$ maximize $A_2$
\citep{RQ36}. In particular $\gamma$ is nearly $180^\circ$, $A$ is
0.01, and $\rho$ is low (0.96 g/cm$^3$). On the other hand $\Theta =
4.33$ for which $f(\Theta)\simeq 0.15$ while the maximum is $\simeq
0.21$. For these reasons we selected those Yarkovsky detections for
which ${\cal S} = |A_2/(A_2)_{exp}|$ was smaller than 1.5. The
selected threshold allows some tolerance as we are scaling only by $D$
without accounting for other quantities such as bulk density, thermal
properties, obliquity, spin rate, and surface roughness.

A high SNR threshold is likely to produce robust detections with
respect to the astrometric data treatment. With lower SNR the
sensitivity to the observation error model increases and detections
become less reliable.
We decided that 3 was a sensible choice for minimum SNR, even if we
analyze detections with smaller SNR in Sec.~\ref{s:low_snr}.

\section{Results}
We applied our 7-dimensional differential corrector to determine the
parameter $A_2$ and the corresponding $da/dt$ for all known NEAs.
After applying the filters discussed in Sec.~\ref{s:filter} we obtain
21 Yarkovsky detections that we consider reliable
(Table~\ref{tab:yarko_tab}).  The reported uncertainties are marginal,
i.e., they fully take into account the correlation between $A_2$ (and
thus $da/dt$) and the orbital elements.

We cross-checked these detections by using two independent software
suites: the JPL Comet and Asteroid Orbit Determination Package and
OrbFit (http://adams.dm.unipi.it/orbfit/)\footnote{OrbFit was used in
  the development version 4.3, currently in beta-testing.}: in all
cases we found agreement at better than the 1-$\sigma$ level.

\begin{center}
\begin{sidewaystable*}[ht]
{\scriptsize
{\hfill{}
  \caption{Semimajor axis, eccentricity, absolute magnitude, physical
    and Yarkovsky parameters, and observational information for
    selected NEAs. The physical quantities for 1999 RQ$_{36}$ are from
    \citet{RQ36}. Golevka's obliquity $\gamma$ is from \citet{hudson},
    Apollo's from \citet{apollo_ito}, Nyx's from \citet{nyx},
    Ra-Shalom's from \citet{ra_shalom}, YORP's from \citet{yorp},
    Geographos' from \citet{geographos}, and Toro's from \v{D}urech
    (personal communication). Diameters $D$ and rotation periods $P$
    are from the European Asteroid Research Node (EARN) database
    (http://earn.dlr.de/nea/). Bond albedo $A$ was computed from
    geometric albedo $p_V$ and slope parameter $G$ (both from the EARN
    database when available) by $A = (0.290 + 0.684 G) p_V$
    \citep{bowell}.}
 \label{tab:yarko_tab}
  \begin{tabular}{@{}lcccccccccccccccc@{}}
    \hline
    \hline
   Asteroid     & $a$ & $e$ & $H$ & $D$ & $P$ & A &$\gamma$ &
    $A_2$ & SNR & ${\cal S}$ & $da/dt$ & $da/dt$ & Diff. & Observed& Radar\\
    & [au] & & & [km] & [hr] & & & [$10^{-15}$ au/d$^2$] &
    & & [$10^{-4}$ au/Myr] &  Nugent et al. & $\sigma$ & arc  &  apparitions\\ 
    \hline
    (101955) 1999 RQ$_{36}$ & 1.13 & 0.20 & 20.6 & 0.49 & 4.29 & 0.01 & 175$^\circ$ & -45.49 $\pm$ 0.23 
    & 197.7 & 1.0 & -18.99 $\pm$ 0.10 & -18.9 $\pm$ 0.2 & 0.4 &
    1999-2012 & 1999, 2005,\\
    &&&&&&&&&&&&&&&  2011\\
    (152563) 1992 BF & 0.91 & 0.27 & 19.6 & 0.51 & NA & 0.02 & NA & -24.01 $\pm$ 1.21 
    & 19.8 & 0.5 & -11.55 $\pm$ 0.58 & -12.84 $\pm$ 1 & 1.1 & 1953-2011 & NA\\
    (6489) Golevka & 2.52 & 0.60 & 19.1 & 0.27 & 6.03 & 0.23 & 137$^\circ$ & -15.88 $\pm$ 1.52 
    & 10.4 & 0.2 & -6.62 $\pm$ 0.64 & -5.74 $\pm$ 0.7 & 0.9 &
    1991-2011 & 1991, 1995,\\
    &&&&&&&&&&&&&&&  2003\\
    2009 BD & 1.01 & 0.04 & 28.2 & NA & NA & NA & NA & -1164.01 $\pm$ 138.76 
    & 8.3 & 0.4 & -493.39 $\pm$ 58.81 & NA & NA & 2009-2011 & NA\\
    (1862) Apollo & 1.47 & 0.56 & 16.0 & 1.40 & 3.06 & 0.12 & 162$^\circ$ & -3.34 $\pm$ 0.52 
    & 6.4 & 0.2 & -1.70 $\pm$ 0.26 & -2.3 $\pm$ 0.2 & 1.8 & 1930-2012 & 1980, 2005\\
    (2062) Aten & 0.97 & 0.18 & 16.9 & 1.30 & 40.77 & NA & NA & -15.41 $\pm$ 2.45 
    & 6.3 & 0.9 & -6.29 $\pm$ 1.10 & -7.5 $\pm$ 2.4 & 0.4 & 1955-2012 & 1995, 2012\\
    (3908) Nyx & 1.93 & 0.46 & 17.3 & 1.00 & 4.43 & NA & 20$^\circ$ & 29.95 $\pm$ 5.39 
    & 5.6 & 1.3 & 11.61 $\pm$ 2.09 & 12.9 $\pm$ 2.7 & 0.4 & 1980-2012 & 1988, 2004\\   
    (2100) Ra-Shalom & 0.83 & 0.44 & 16.1 & 2.24 & 19.80 & 0.05 & 162$^\circ$ & -10.97 $\pm$ 2.25 
    & 4.9 & 1.1 & -6.31 $\pm$ 1.30 & -5.4 $\pm$ 1.5 & 0.5 & 1975-2012
    & 1981, 1984,\\
    &&&&&&&&&&&&&&&  2000, 2003\\
    (10302) 1989 ML & 1.27 & 0.14 & 19.4 & 0.25 & 19 & NA & NA & 90.48 $\pm$ 16.38
    & 4.6 & 1.0 & 34.71 $\pm$ 6.28 & 35.3 $\pm$ 7.1 & 0.0 & 1989-2012 & NA\\
    1999 MN & 0.67 & 0.67 & 21.4 & NA & 5.49 & NA & NA & 50.79 $\pm$ 11.39 
    & 4.5 & 0.4 & 47.12 $\pm$ 10.56 & NA & NA & 1999-2010 & NA\\
    (2340) Hathor & 0.84 & 0.45 & 19.9 & 0.30 & NA & 0.24 & NA & -24.71 $\pm$ 5.66 
    & 4.4 & 0.3 & -14.33 $\pm$ 3.28 & -14.5 $\pm$ 3.5 & 0.0 & 1976-2012 & NA\\
    (6037) 1988 EG & 1.27 & 0.50 & 18.7 & 0.40 & 2.76 & NA & NA & -32.66 $\pm$ 8.19 
    & 4.0 & 0.6 & -16.39 $\pm$ 4.11 & -14.2 $\pm$ 4.3 & 0.4 & 1988-2011 & 1988\\
    (37655) Illapa & 1.48 & 0.75 & 17.8 & NA & 2.66 & NA & NA & -13.99 $\pm$ 3.65 
    & 3.8 & 0.6 & -11.27 $\pm$ 2.94 & NA & NA & 1994-2012 & 2012\\
    (85953) 1999 FK$_{21}$ & 0.74 & 0.70 & 18.0 & 0.59 & NA & NA & NA & -10.62 $\pm$ 2.33 
    & 3.7 & 0.3 & -10.38 $\pm$ 2.28 & -10.44 $\pm$ 1.5 & 0.0 & 1971-2012 & NA\\
    (1685) Toro & 1.37 & 0.44 & 14.3 & 3.00 & 10.20 & 0.05 & 144$^\circ$ & -2.83 $\pm$ 0.77 
    & 3.7 & 0.4 & -1.27 $\pm$ 0.34 & -1.4 $\pm$ 0.7 & 0.2 & 1948-2012 & 1980,
    1988\\
    &&&&&&&&&&&&&&&  2012\\ 
    2005 ES$_{70}$ & 0.76 & 0.39 & 23.6 &  NA & NA & NA & NA & -97.23 $\pm$ 29.27 
    & 3.3 & 0.3 & -55.58 $\pm$ 16.73 & NA & NA & 2005-2011 & NA\\
    (54509) YORP & 1.00 & 0.23 & 22.6 & 0.10 & 0.20 & NA & 178$^\circ$ & -74.93 $\pm$ 23.68 
    & 3.2 & 0.3 & -33.60 $\pm$ 10.61 & -35.63 $\pm$ 10.5 & 0.1 &
    2000-2005 & 2001, 2004\\
    (283457) 2001 MQ$_3$ & 2.23 & 0.46 & 18.9 & NA & NA & NA & NA & -44.63 $\pm$ 14.33 
    & 3.1 & 1.1 & -16.02 $\pm$ 5.14 & NA & NA & 1951-2011 & NA\\
    (1620) Geographos & 1.25 & 0.34 & 15.2 & 2.47 & 5.22 & NA & 153$^\circ$ & -4.25 $\pm$ 1.40 
    & 3.0 & 0.5 & -1.82 $\pm$ 0.60 & -2.5 $\pm$ 0.6 & 0.8 & 1951-2012 & 1983, 1994\\
    (65679) 1989 UQ & 0.91 & 0.26 & 19.3 & 0.73 & 7.73 & NA & NA & -36.66 $\pm$ 12.23 
    & 3.0 & 1.2 & -17.51 $\pm$ 5.84 & NA & NA & 1954-2011 & NA\\
    (162004) 1991 VE & 0.89 & 0.66 & 18.0 & NA & NA & NA & NA & 18.30 $\pm$ 6.13 
    & 3.0 & 0.7 & 14.75 $\pm$ 4.94 & NA & NA & 1954-2012 & NA\\
\hline
\end{tabular}}
\hfill{}}
\end{sidewaystable*}
\end{center}

\subsection{2009 BD}
Asteroid 2009 BD is very small and to fit its observational dataset it
is necessary to include solar radiation pressure
\citep{2009BD}. Therefore, we also included in the model a radial
acceleration $a_r = A_1/r^2$.  Along with the tabulated value of
$A_2$, we obtained $A_1 = (62.05 \pm 8.85) \times 10^{-12}$
au/d$^2$. This results in an area to mass ratio $A/M = (2.72 \pm 0.39)
\times 10^{-4}$ m$^3$/kg, which is consistent at the 1-$\sigma$ level
with the value reported by \citet{2009BD}, i.e., $(2.97 \pm 0.33)
\times 10^{-4}$ m$^3$/kg.

\subsection{Comparison with other published results}\label{s:comparison}
The first three objects of Table~\ref{tab:yarko_tab} are the already
known cases of Golevka, 1992 BF, and 1999 RQ$_{36}$.  While for 1999
RQ$_{36}$ there is a perfect match between our result and the one in
\citet{RQ36}, for Golevka and 1992 BF the values are different from
\citet{golevka} and \citet{1992BF}, respectively. However, this can be
easily explained by the availability of new astronomy and the fact
that the present paper adopted the debiasing and weighting scheme by
\citet{cbm10}, which was not available at the time of the earlier
publications.

As already mentioned, \citet{nugent} performed a search similar to the
one presented in this paper and found 54 NEAs with a measurable
semimajor axis drift. The main differences are the following:
\begin{itemize}
\item They selected only numbered objects, while we included all known
  NEAs.
\item Their observation dataset was slightly different as they used
  observations until 2012 January 31, while we have data until 2012
  October 31. This difference does not really matter for optical data,
  but it does for radar data, e.g., for Aten and Toro. Moreover, they
  did not use single apparition radar, while we did as we think they
  represent an important constraint.
\item They solved for constant $da/dt$ while we used constant $A_2$
  and then convert to $da/dt$. These techniques are equivalent when
  the semimajor axis and eccentricity are constant, but there could be
  differences as we cannot assume $da/dt$ constant for objects
  experiencing deep planetary close approaches.
\item They searched for the best-fit $da/dt$ by means of the golden
  section algorithm, i.e., they computed the RMS of the residuals
  corresponding to the best fitting orbital elements for fixed values
  of $da/dt$, then the minimum is obtained by interpolation. In this
  paper we used a full 7-dimensional differential corrector. The two
  methods should be equivalent.
\item They used 1 as lower bound for SNR, while we use 3 that gives
  detections more robust against changes in the observation weighting.
  Also, they used the ``sensitivity'' parameter, i.e., a metric to
  measure the sensitivity of a dataset to the presence of a semimajor
  axis drift. We did not use such a metric as we think that an
  SNR$\geq 3$
  is already a good metric in that respect.
\item We kept only those objects for which the measured orbital drift
  can be related to the Yarkovsky perturbation presuming that
  inconsistencies stem from astrometric errors, while they also
  considered the possibility of other nongravitational effects such as a
  loss of mass.
\end{itemize}

Table~\ref{tab:yarko_tab} contains a comparison between our orbital
drifts and the ones reported by \citet{nugent}. 2009 BD, 1999 MN, 1999
FA, and 2005 ES$_{70}$ are present only in our list as they are not
numbered, while 2001 MQ$_3$, 1989 UQ, and 1991 VE are eliminated by
their filters. It is worth pointing out that also 1999 RQ$_{36}$, 1992
BF, Golevka, and YORP have been filtered out by \citet{nugent}
criteria, even though they report the corresponding detections for a
comparison with \cite{chesley_list}.

Among the cases that \citet{nugent} report with $\text{SNR}>3$ we
neglected the following three:
\begin{itemize}
\item (1036) Ganymed for which we found $A_2 = (-16.54 \pm 4.35)
  \times 10^{-15}$ au/d$^2$, corresponding to $da/dt = (-6.06 \pm
  1.59) \times 10^{-4}$ au/Myr. However, the nominal $A_2$ is 28 times
  larger than $(A_2)_{exp}$, so we marked this detection as
  spurious. As Ganymed observations go back to 1924, this unreliable
  detection might be due to bad quality astrometry.
\item (4197) 1982 TA for which we used the radar apparition of 1996,
  which reduced the SNR below 1. For this object we found $A_2 = (5.61
  \pm 14.26) \times 10^{-15}$ au/d$^2$, corresponding to $da/dt =
  (3.88 \pm 9.86) \times 10^{-4}$ au/Myr. For comparison
  \citet{nugent} report $da/dt = (30.9 \pm 9.2) \times 10^{-4}$
  au/Myr.
\item (154330) 2002 VX$_4$ for which we found $A_2 = 102.36 \pm 36.34)
  \times 10^{-15}$ au/d$^2$, corresponding to $da/dt = (43.10 \pm
  15.25) \times 10^{-4}$ au/Myr. Again, the value of $A_2$ was $\sim$
  4 times larger than $(A_2)_{exp}$, so also this detection was marked
  as spurious.
\end{itemize}

Besides the differences outlined above, there is an overall agreement
for the common cases. As a matter of fact by computing the $\sigma$ of
the difference, i.e., $\sigma^2 = {\sigma_1^2 + \sigma_2^2}$, there
are only two cases that are not consistent at the 1-$\sigma$ level:
\begin{itemize}
\item 1992 BF, for which we used remeasured observations
  \citep{1992BF} and changed the weights (see Sec.~\ref{s:cbm}) for
  the 1953 apparition. The remeasurements are not available in the MPC
  database.
\item Apollo, for which we applied a suitable manual weighting as
  described in Sec.~\ref{s:cbm}.
\end{itemize}


\subsection{Lower SNR and other less reliable detections}\label{s:low_snr}
\begin{figure*}[ht!]
  \centering
  \includegraphics[width=8cm]{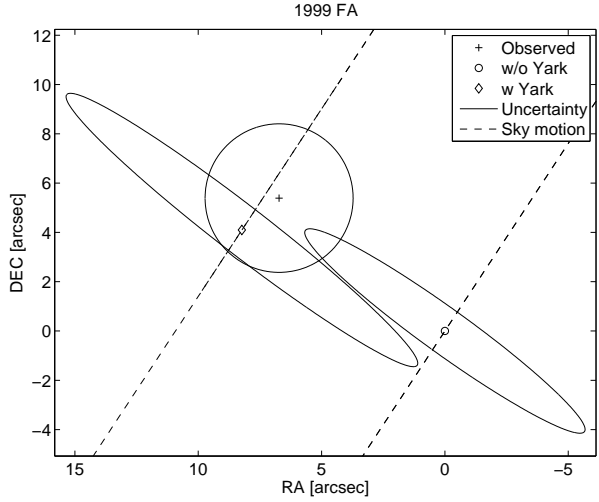}
  \caption{Same of Fig.~\ref{fig:precov} for asteroid 1999 FA. The
    uncertainty ellipses for predicted positions correspond to
    3-$\sigma$ levels, while the uncertainty for the observaed
    position correspond to 3''.}
  \label{fig:1999FA}
\end{figure*} 

\begin{figure*}[ht!]
 \centering
 \includegraphics[width=8cm]{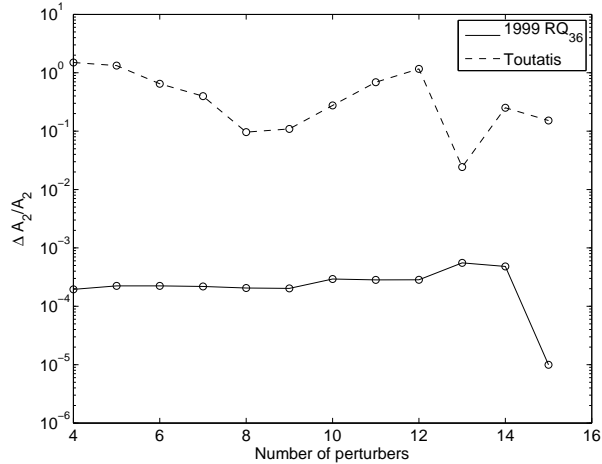}
 \caption{Relative displacement of $A_2$ with respect to the nominal
   value as a function of the number of perturbers for 1999 RQ$_{36}$
   and Toutatis.}
 \label{fig:perturbers}
\end{figure*}
Table~\ref{tab:yarko_tab_b} contains detections that we rate as
nonspurious on the basis of the $S$ ratio between expected and measured
value, down to SNR $=$ 2 plus the following remarkable cases:
\begin{itemize}
\item 1999 FA, which has 1 isolated observation in 1978 from Siding
  Spring Observatory. By performing the same analysis of
  Sec.~\ref{s:precov} (see Fig.~\ref{fig:1999FA}) we see no clear
  improvement due to Yarkovsky as the observation is 3.5--4 $\sigma$
  away from the prediction either way. In this case a clock error may
  bring the Yarkovsky solution close to the observation. As this
  detection depends on a lone, isolated observation we would rather be
  cautious and consider this detection reliable only if the 1978
  observation is remeasured.
\item Eros, which looks like a reliable detection as ${\cal S} =
  0.75$. However, the obliquity is known to be 89$^\circ$ \citep{eros}
  and then we enter the regime where the seasonal component of the
  Yarkovsky effect is dominant. As the seasonal component is typically
  10 times smaller then the diurnal one \citep{vok_2000} we mark this
  detection as spurious. This points to possible bad astrometric
  treatment, especially for historical observations dating back to
  1893.
\item Toutatis, which enters the Main Belt region because of the 4.12
  au aphelion and the low inclination (0.44$^\circ$). Therefore, it is
  important to account for the uncertainty in the masses of the
  perturbing asteroids. By taking into account the uncertainty of the
  perturbing asteroid masses the actual uncertainty in $A_2$ increases
  by 11\% with a commensurable drop in SNR.
  Figure~\ref{fig:perturbers} shows the evolution of $A_2$ as a
  function of the number of perturbers for asteroids 1999 RQ$_{36}$
  and Toutatis. While for 1999 RQ$_{36}$ (aphelion 1.36 au) we reach
  convergence with just four perturbers, for Toutatis we have a quite
  irregular behavior suggesting that we may need to include more
  perturbers.
\item 1994 XL$_1$, whose observations in 1994 from Siding Spring
  Observatory show high residual so we relaxed weights to 3''.
\item 2005 QC$_5$ and 2000 NL$_{10}$ have been included despite the
  low SNR. Similarly to the cases described in Sec.~\ref{s:precov}, we
  applied weights 3'' to precovery observations and this data
  treatment weakened the Yarkovsky detection. However, we think that
  remeasuring the precovery observations may lead to reliable
  Yarkovsky detections for these objects.
\item 1950 DA, for which the Yarkovsky effect plays an important role
  for impact predictions, e.g., see Sec.~\ref{s:impact}.
\item Icarus, for which the low values of both the nominal value and
  the uncertainty put a strong contraint on $A_2$. In particular, it
  is worth pointing out that the constraint on $da/dt$ is consistent
  with \citet[Fig.~5]{vok_2000} where $|da/dt|<3 \times 10^{-4}$
  au/Myr.
\end{itemize}
There are other objects with an even lower SNR for which the Yarkovsky
signal might be revealed if precovery observations were remeasured:
(11284) Belenus, (66400) 1999 LT$_7$, (4688) 1980 WF, (67399) 2000
PJ$_6$, (267759) 2003 MC$_7$, and (88710) 2001 SL$_9$.

Though these detections have to be considered less reliable, some of
them may be good candidates for becoming stronger detections in the
future if high quality astrometry is obtained, e.g., by radar or GAIA
\citep{gaia}.

\begin{center}
\begin{sidewaystable*}[ht]
{\scriptsize
{\hfill{}
  \caption{Same as Table~\ref{tab:yarko_tab} for less reliable
    detections. Icarus' obliquity is from \citet{icarus}, Eros' from
    \citet{eros}, Nereus' from \citet{Nereus}, and Betulia's from
    \citet{nyx}. All the other physical quantities are from the EARN
    database.}
 \label{tab:yarko_tab_b}
 \begin{tabular}{@{}lcccccccccccccccc@{}}
    \hline
    \hline
    Asteroid & $a$ & $e$ & $H$ & $D$ & $P$ & A &$\gamma$ &
    $A_2$ & SNR & ${\cal S}$ & $da/dt$ & $da/dt$ & Diff. & Observed& Radar\\
    & [au] & & & [km] & [hr] & & & [$10^{-15}$ au/d$^2$] &
    & & [$10^{-4}$ au/Myr] &  Nugent et al. & $\sigma$ &  arc  &  apparitions\\ 
    \hline
    1999 FA & 1.07 & 0.13 & 20.6 & 0.30 & 10.09 & NA & NA & -98.69 $\pm$ 6.15 
    & 3.2 & 1.3 & -41.07 $\pm$ 13.02 & NA & NA & 1978-2008 & NA\\
    (433) Eros & 1.46 & 0.22 & 10.8 & 23.30 & 5.27 & 0.10 & 89$^\circ$ & -0.72 $\pm$ 0.23 
    & 3.1 & 0.7 & -0.26 $\pm$ 0.09 & -0.3 $\pm$ 0.2 & 0.2 &
    1893-2012 & 1975, 1988\\
    (4660) Nereus & 1.49 & 0.36 & 18.1 & 0.34 & 15.16 & NA & 11$^\circ$ & 28.58 $\pm$ 11.72 
    & 2.4 & 0.4 & 11.43 $\pm$ 4.69 & 10.9 $\pm$ 4.8 & 0.1 &
    1981-2012 & 2001, 2002\\
    2007 PB$_8$ & 0.88 & 0.45 & 21.2 & NA & NA & NA & NA & -156.01 $\pm$ 66.45 
    & 2.3 & 1.4 & -88.05 $\pm$ 37.50 & NA & NA & 2002-2012 & NA\\
    2004 BG$_{41}$ & 2.52 & 0.61 & 24.4 & NA & NA & NA & NA & -598.99 $\pm$ 255.87 
    & 2.3 & 1.2 & -256.03 $\pm$ 109.37 & NA & NA & 2004-2011 & NA\\
    (4179) Toutatis & 2.53 & 0.63 & 15.1 & 2.80 & 176.00 & 0.05 & NA & -3.32 $\pm$ 1.43 
    & 2.3 & 0.4 & -1.49 $\pm$ 0.63 & -5.0 $\pm$ 0.6 & 4.1 &
    1976-2012 & 1992, 1996, \\
    &&&&&&&&&&&&&&&  2000, 2004,\\
    &&&&&&&&&&&&&&& 2008\\
    (138911) 2001 AE$_2$ & 1.35 & 0.08 & 19.0 & 0.35 & NA & NA & NA & -62.23 $\pm$ 28.84 
    & 2.2 & 1.0 & -22.90 $\pm$ 10.61 & -22.9 $\pm$ 11.2 & 0.0 &
    1984-2012 & NA\\
    (326290) 1998 HE$_3$ & 0.88 & 0.44 & 21.7 & 0.10 & NA & NA & NA & -70.57 $\pm$ 33.01 
    & 2.1 & 0.3 & -39.68 $\pm$ 18.56 & NA & NA &
    1993-2012 & 2012\\
    (3361) Orpheus & 1.21 & 0.32 & 18.9 & 0.50 & 3.51 & NA & NA & 12.07 $\pm$ 5.67 
    & 2.1 & 0.3 & 5.20 $\pm$ 2.44 & 5.7 $\pm$ 2.5 & 0.1 &
    1982-2010 & NA\\
    (154590) 2003 MA$_3$ & 1.11 & 0.40 & 21.7 & NA & NA & NA & NA & -95.60 $\pm$ 45.11 
    & 2.1 & 0.7 & -46.06 $\pm$ 21.74 & NA & NA &
    1998-2012 & NA\\
    1994 XL$_1$ & 0.67 & 0.53 & 20.8 & NA & NA & NA & NA & -31.22 $\pm$ 15.32 
    & 2.0 & 0.3 & -22.38 $\pm$ 10.98 & NA & NA & 1994-2011 & NA\\
    (1580) Betulia & 2.19 & 0.49 & 14.3 & 4.57 & 6.13 & 0.03 & 117$^\circ$ & -4.65 $\pm$ 2.31 
    & 2.0 & 0.9 & -1.75 $\pm$ 0.87 & -1.3 $\pm$ 0.9 & 0.4 &
    1950-2012 & 1976, 1989,\\
    &&&&&&&&&&&&&&& 2002\\
    (29075) 1950 DA & 1.70 & 0.51 & 17.0 & 1.4 & 2.12 & 0.08 & NA & -5.00 $\pm$ 3.47 
    & 1.4 & 0.3 & -2.20 $\pm$ 1.52 & NA & NA &
    1950-2012 & 2001\\
    (250680) 2005 QC$_5$ & 0.89 & 0.36 & 19.7 & NA & NA & NA & NA & 26.44 $\pm$ 22.01 
    & 1.2 & 0.5 & 13.70 $\pm$ 11.40 & NA & NA &
    1978-2011 & NA\\
    (105140) 2000 NL$_{10}$ & 0.91 & 0.82 & 15.5 & 1.72 & 6.93 & NA & NA & -11.78 $\pm$ 10.49 
    & 1.1 & 0.9 &  -15.73 $\pm$ 14.02 & NA & NA &
    1951-2012 & NA\\
    (1566) Icarus & 1.08 & 0.83 & 16.7 & 1.30 & 2.27 & 0.04 & 76$^\circ$ & -0.66 $\pm$ 1.39 
    & 0.5 & 0.1 & -0.86 $\pm$ 1.80 & -3.2 $\pm$ 2.0 & 0.9 &
    1949-2012 & 1968, 1996\\
    \hline
\end{tabular}
\hfill{}}}
\end{sidewaystable*}
\end{center}

\section{Discussion}

\subsection{Connection with NEA feeding mechanisms}
The diurnal Yarkovsky effect produces a semimajor axis drift
proportional to $\cos\gamma$ \citep{vok_2000}. As the diurnal term is
typically the dominant one, the sign of $da/dt$ can be related to the
asteroid spin orientation, i.e., a negative $da/dt$ corresponds to a
retrograde rotator while a positive $da/dt$ corresponds to a prograde
rotator. This conclusion is supported by the eight known obliquities for the
asteroids in the sample that are listed in Table~\ref{tab:yarko_tab}:
in all cases the spin axis obliquity is consistent with the sign of
$da/dt$.

We can now use this interpretation and our solution for the Yarkovsky
semimajor axis drift values for NEAs in the following
way. Table~\ref{tab:yarko_tab} contains four prograde rotators and
seventeen retrograde rotators. This excess of retrograde rotators can
be explained by the nature of resonance feeding into the inner Solar
System \citep{population}. Most of the primary NEA source regions
(e.g., the 3:1 resonance, JFCs, Outer Belt, etc.) allow main belt
asteroids to enter by drifting either inwards or outwards, but the
$\nu_6$ resonance is at the inner edge of the main belt and so
asteroids can generally enter only by inwards drift, i.e., with
retrograde rotation. \citet{population} report that 37\% of NEAs
arrive via $\nu_6$ resonance. \citet{laspina} point out that this
implies 37\% of NEAs have retrograde spin (via $\nu_6$), plus half of
the complement (via other pathways). Thus, the retrograde fraction
should be $0.37 + 0.5 \times 0.63 = 0.69$, while \citet{laspina}
report 67\% retrograde for their sample, which is dominated by large
NEAs.

Table~\ref{tab:yarko_tab} contains 81\% retrograde rotators, which is
larger than 69\% and then appears to be inconsistent with the
theory. However, the asteroid sample of Table~\ref{tab:yarko_tab} is
based on measured Yarkovsky mobility, and so is dominated by small
PHAs. For this population, 55\% are expected to arrive from $\nu_6$
according to the \citet{population} NEO model (Bottke, personal
communication), a much greater fraction than for NEAs as a whole. The
same arithmetic used for NEAs gives 0.55+0.5$\times$0.45=0.78, which
matches very well the $81\% \pm 8\%$ value of
Table~\ref{tab:yarko_tab}.

\begin{figure*}
  \centering
  \includegraphics[width=6.8cm]{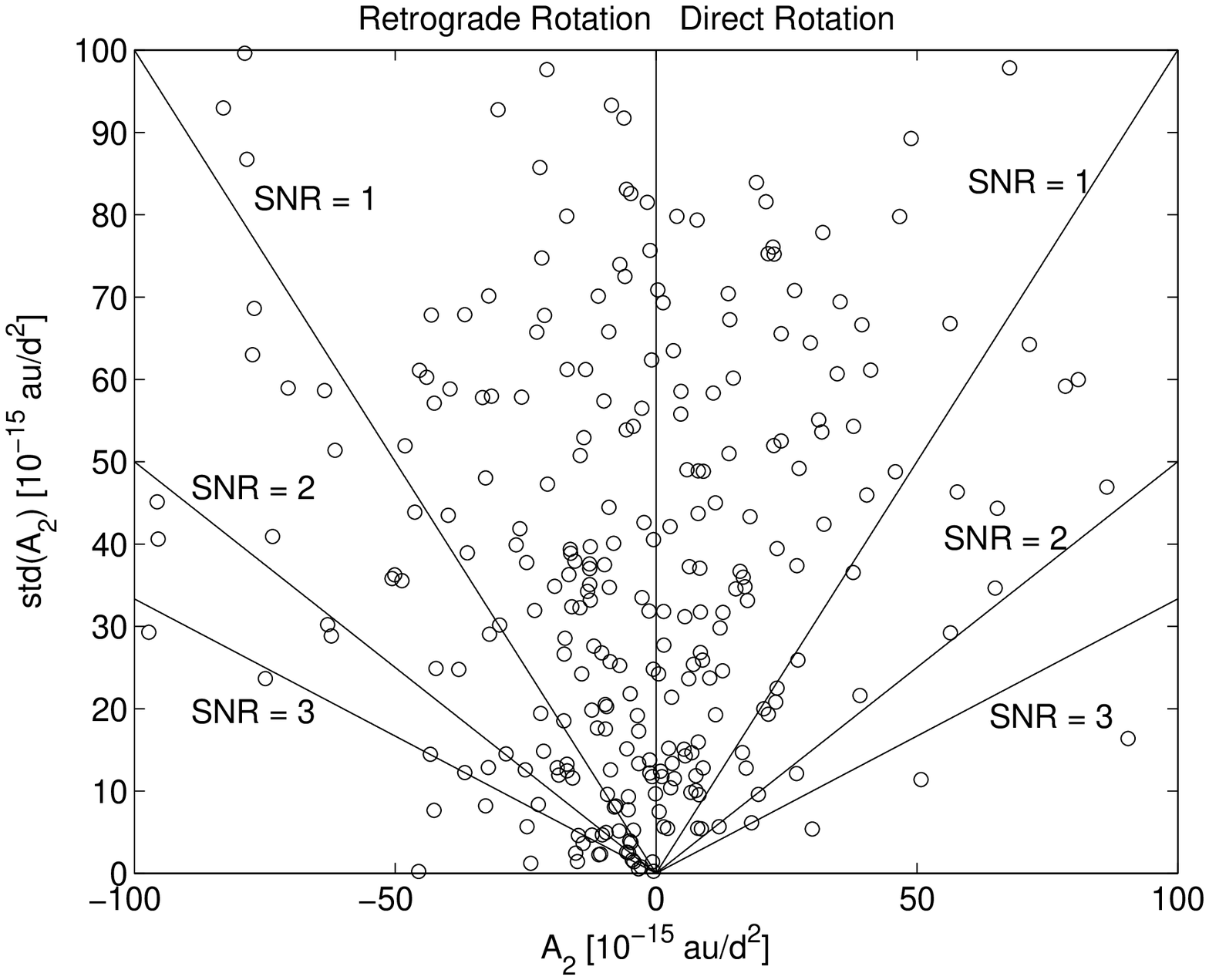}\includegraphics[width=6.8cm]{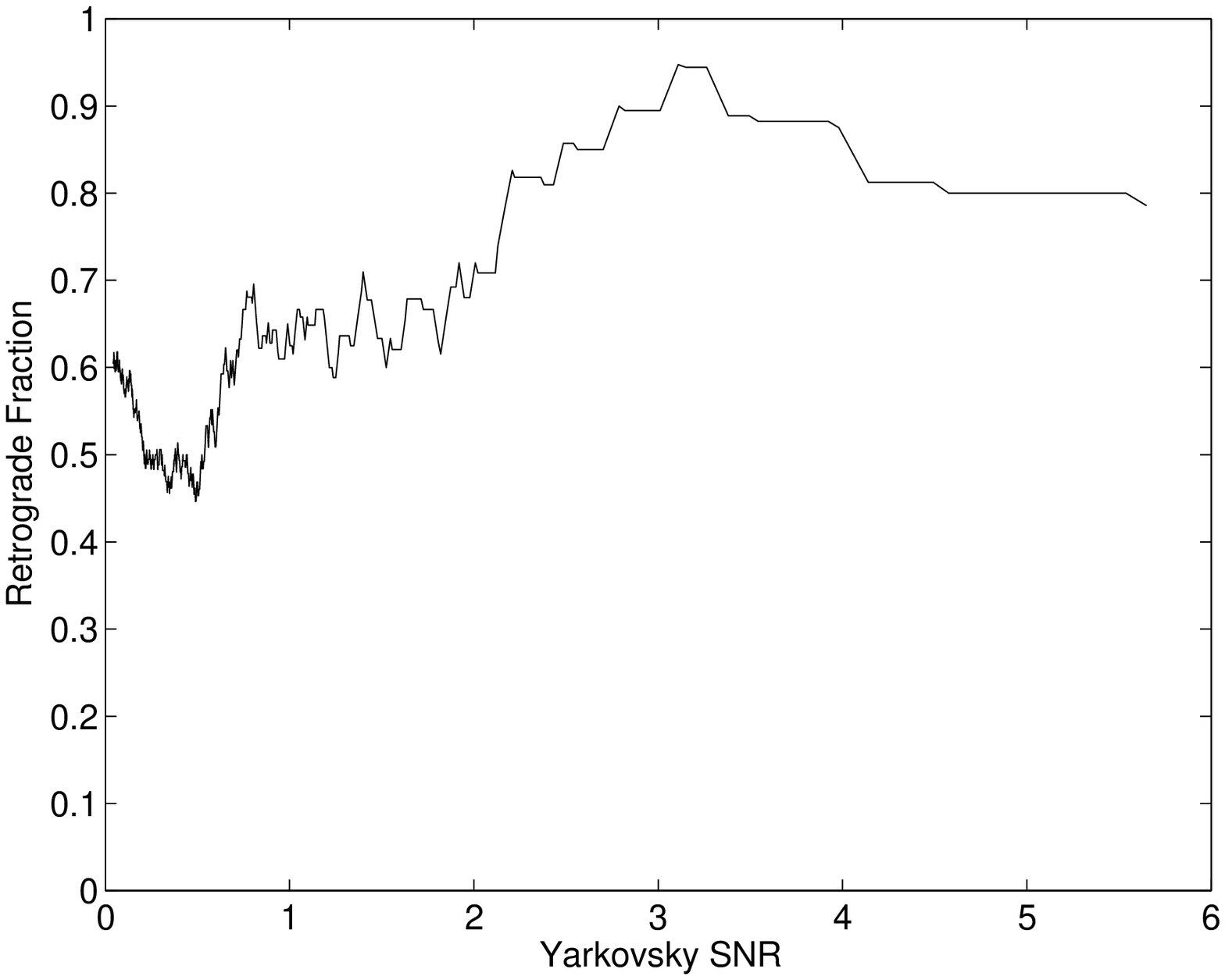}
  \caption{Left panel: distribution of $A_2$ and its
    uncertainty. Right side of plot is prograde rotation and left side
    is retrograde rotation. Right panel: running box mean in SNR for
    the fraction of asteroids with indication of retrograde
    spin. Cases with $\text{SNR}>2$ show a consistent 4:1 retrograde
    ratio, cases with little or no signal are split 50--50.}
  \label{fig:snr}
\end{figure*}
To assess the behavior of the fraction of retrograde rotators as a
function of the SNR, we took all of the objects with ${\cal S}<1.5$.
The left panel of Fig.~\ref{fig:snr} shows the distribution of $A_2$
and its uncertainty. The excess of retrograde rotators is clearly
visible for $\text{SNR}>3$ and also between 2 and 3, where small PHAs
dominate. For lower SNR we have a more uniform distribution. The right
panel of Fig.~\ref{fig:snr} is a running mean of the fraction of
retrograde rotators as a function of the SNR. For $\text{SNR}<1$ we
are in a noise dominated regime for which we have a rough 50\%
fraction of retrograde rotators, for $1< \text{SNR} <2$ we have a
transition from noise-dominated to signal dominated, and for
$\text{SNR}>2$ we have a signal dominated regime with around 80\%
retrograde rotators.

\begin{figure*}
  \centering
  \includegraphics[width=8cm]{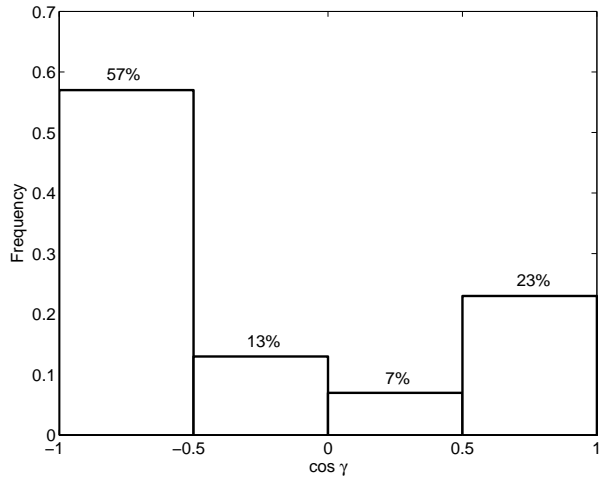}
  \caption{Inferred obliquity distribution for objects with
    $\text{SNR}>1$.}
  \label{fig:obliquity}
\end{figure*}

We can also try to use the detected values to infer the obliquity
distribution. From Eq.~\eqref{eq:A2} we have that
\begin{equation}
A_2 \propto \frac{\cos\gamma}{D} .
\end{equation}
and so we can estimate $\gamma$ by using the either known or estimated
by Eq.~\eqref{eq:diameter} diameter and using 1999 RQ$_{36}$. 
By taking all the detections with $\text{SNR}>1$ and assuming a fixed $\rho =
1500$ kg/m$^3$ we obtain the distribution of Fig.~\ref{fig:obliquity},
where the cases with $|\cos\gamma|>1$ have been placed in the extreme
obliquity bins.
Despite the low number of bins, we can see the excess of retrograde
rotators and the abundance of objects with an extreme obliquity, as
expected from the YORP effect \citep{capek_yorp}. While this
distribution should be considered only approximate due to the numerous
assumptions (e.g., neglecting dependence on bulk density, shape and
thermal properties) we consider it to be a significant improvement
over what is otherwise known. However, we find it interesting that it
appears to be consistent with the observed obliquity distribution of
the NEAs \citep{laspina}.

\subsection{Spurious detections}
\begin{figure*}
  \centering
  \includegraphics[width=6.8cm]{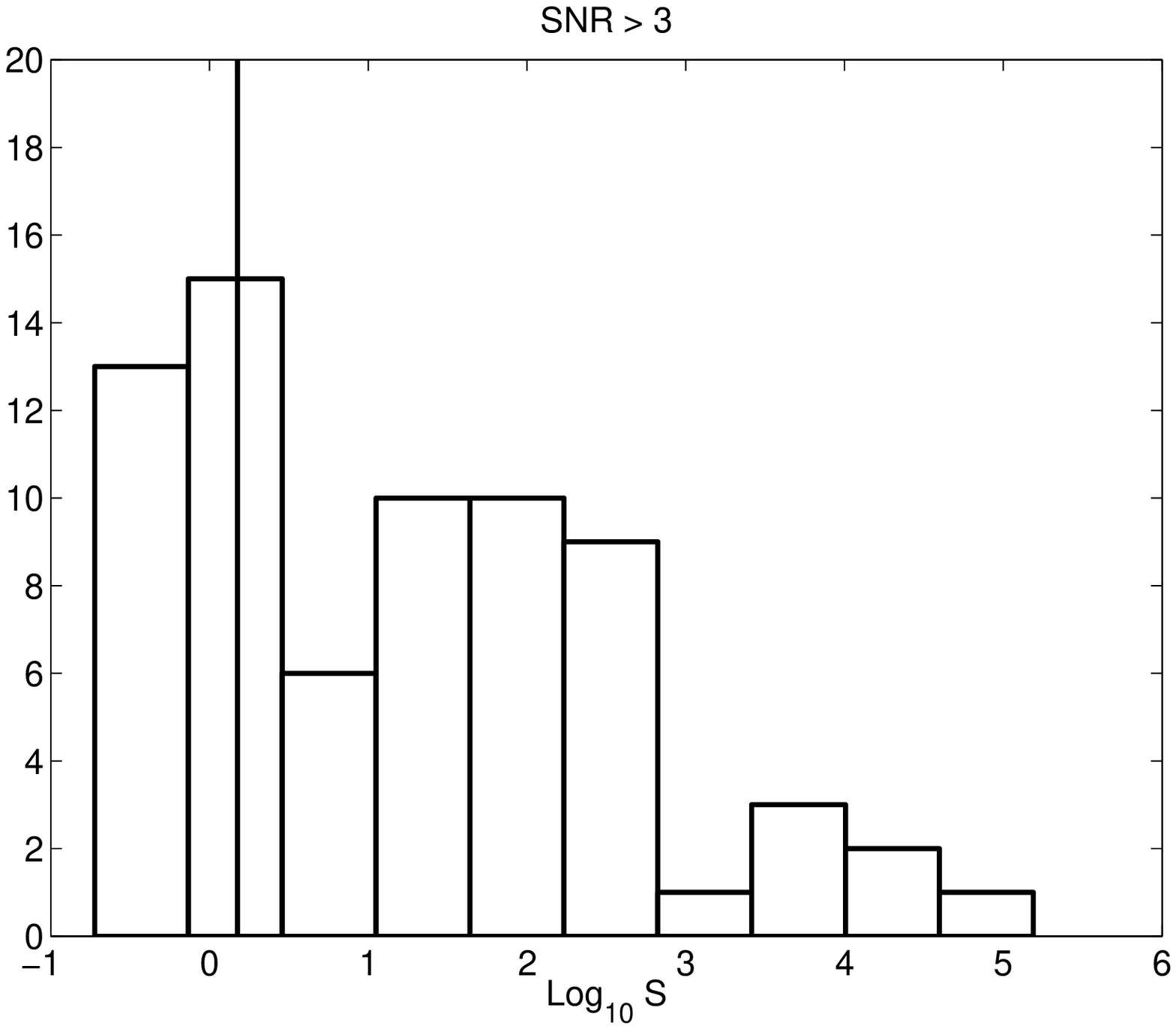}\includegraphics[width=6.8cm]{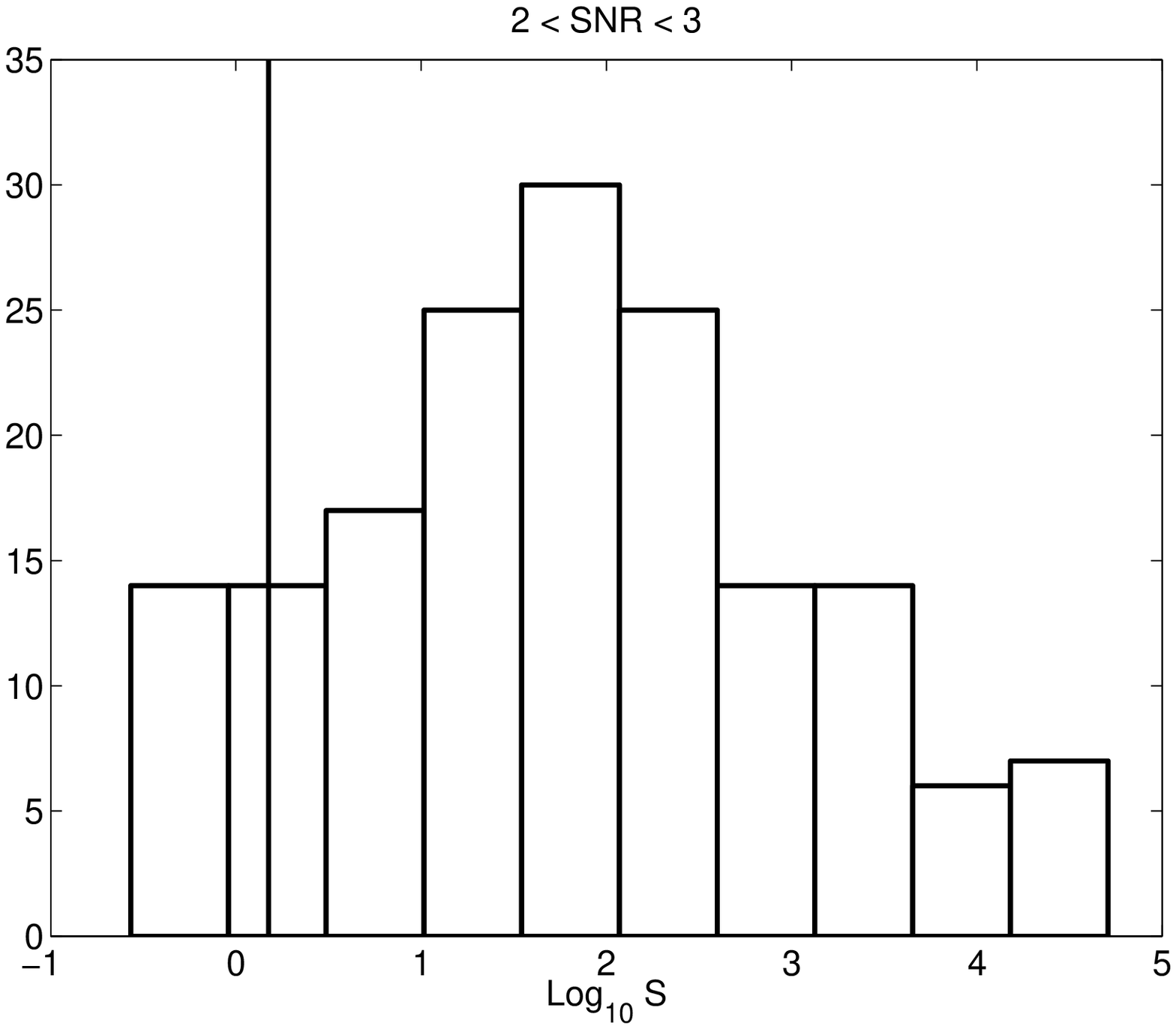}
  \caption{Histogram of ${\cal S}$ for different intervals of SNR. The
    vertical lines correspond to the selected threshold 1.5.}
  \label{fig:ratio}
\end{figure*}

Our search for Yarkovsky signal produced a large number of spurious
detections, i.e., semimajor axis drifts far larger than the Yarkovsky
effect would cause.  Figure~\ref{fig:ratio} contains the histograms of
${\cal S}$ for different SNR intervals.  For $\text{SNR} > 3$ we have
67\% spurious detections, for $2 < \text{SNR} < 3$ we have 88\%.

It it worth trying to understand the reason of these spurious
solutions. We think there are two possible causes:
\begin{itemize}
\item Bad astrometry treatment: as discussed in Sec.~\ref{s:cbm} and
  Sec.~\ref{s:precov}, non-CCD observations may contain errors that
  are difficult to model. If an ad hoc weighting is not used we may
  have misleading results. Indeed, spot checking of such cases
  generally confirmed isolated astrometry as source of spurious
  detections.
\item Incompleteness or inconsistency in the dynamical model: the
  formulation proposed in Sec.~\ref{s:yarko} is a simplified model of
  the Yarkovsky perturbation that might be poor in some cases. A more
  sophisticated formulation would require a rather complete physical
  characterization that is typically unavailable and thus cannot be
  used for a comprehensive search as done in this paper. Moreover, as
  discussed in Sec.~\ref{s:low_snr} for Toutatis, we may need to
  include more perturbing asteroids (and the uncertainty in their
  masses) in the model. Finally, we cannot rule out the possibility of
  nongravitational perturbations different from Yarkovsky.
\end{itemize}

\subsection{Constraining physical quantities}
\begin{figure*}[ht]
  \centering
  \includegraphics[width=8cm]{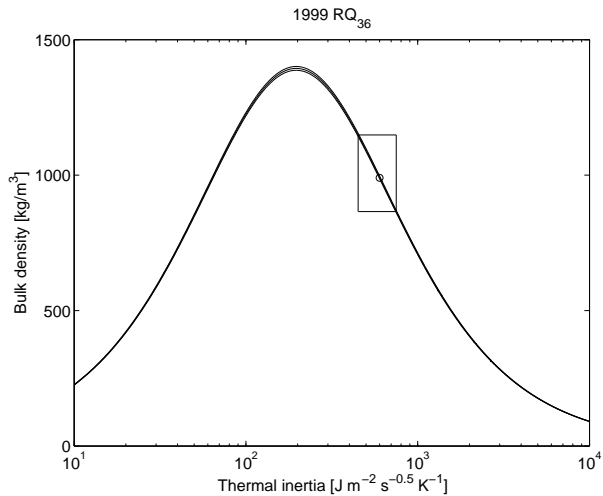}
  \caption{Density as a function of thermal inertia for asteroid 1999
    RQ$_{36}$. The three lines corresponding to the nominal value and
    the 1-$\sigma$ levels are extremely narrow because of the high
    SNR. The circle corresponding to the nominal observed value of
    $\Gamma$ and thus inferred nominal value of $\rho$ \citep{RQ36}
    matches the plotted curves. The enclosed region corresponds to the
    1-$\sigma$ interval for $\Gamma$ \citep{emery} and indicates how
    it maps onto the 1-$\sigma$ limits of $\rho$.}
  \label{fig:rq36}
\end{figure*}

\begin{figure*}[ht]
  \centering
  \includegraphics[width=6.8cm]{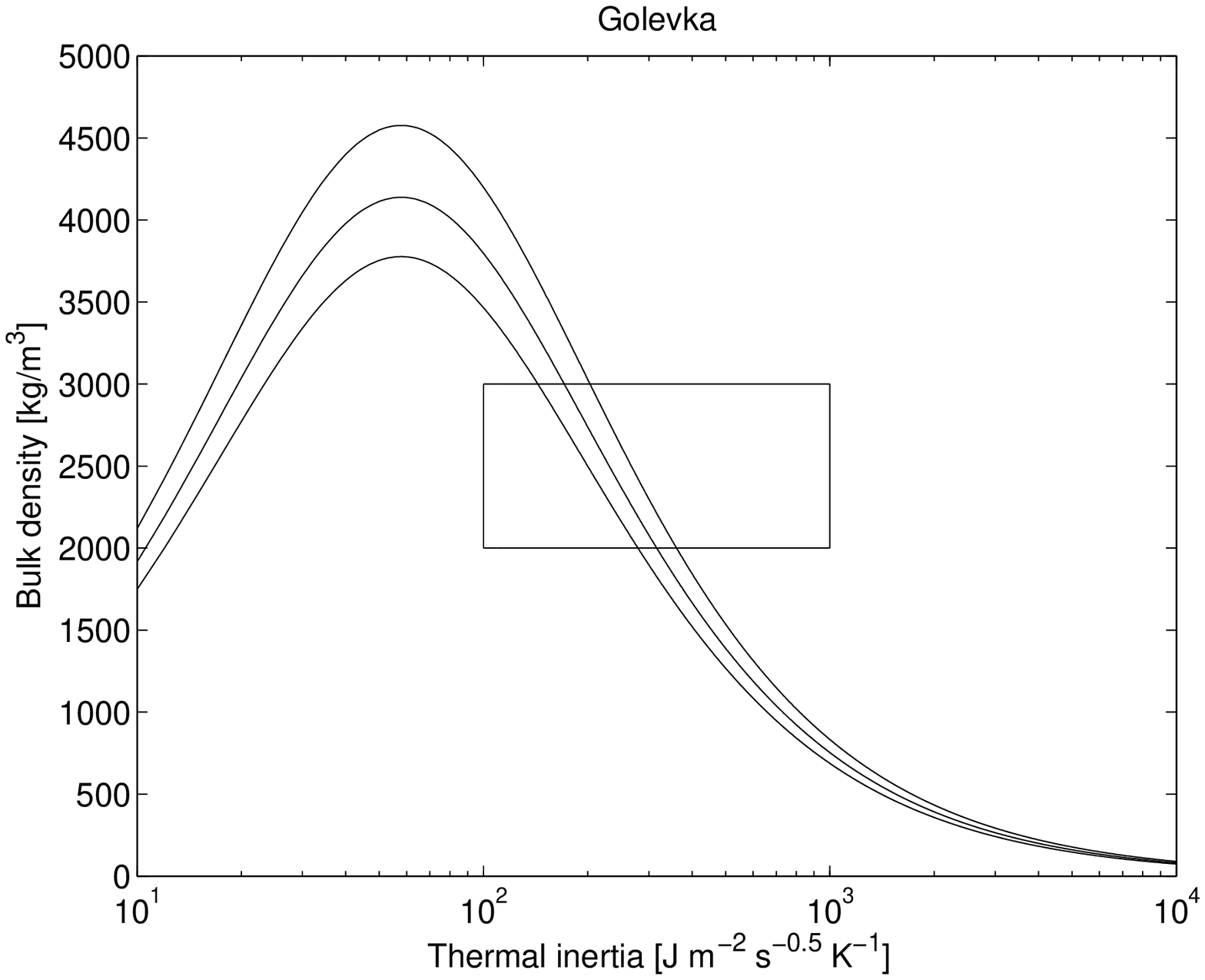}\includegraphics[width=6.8cm]{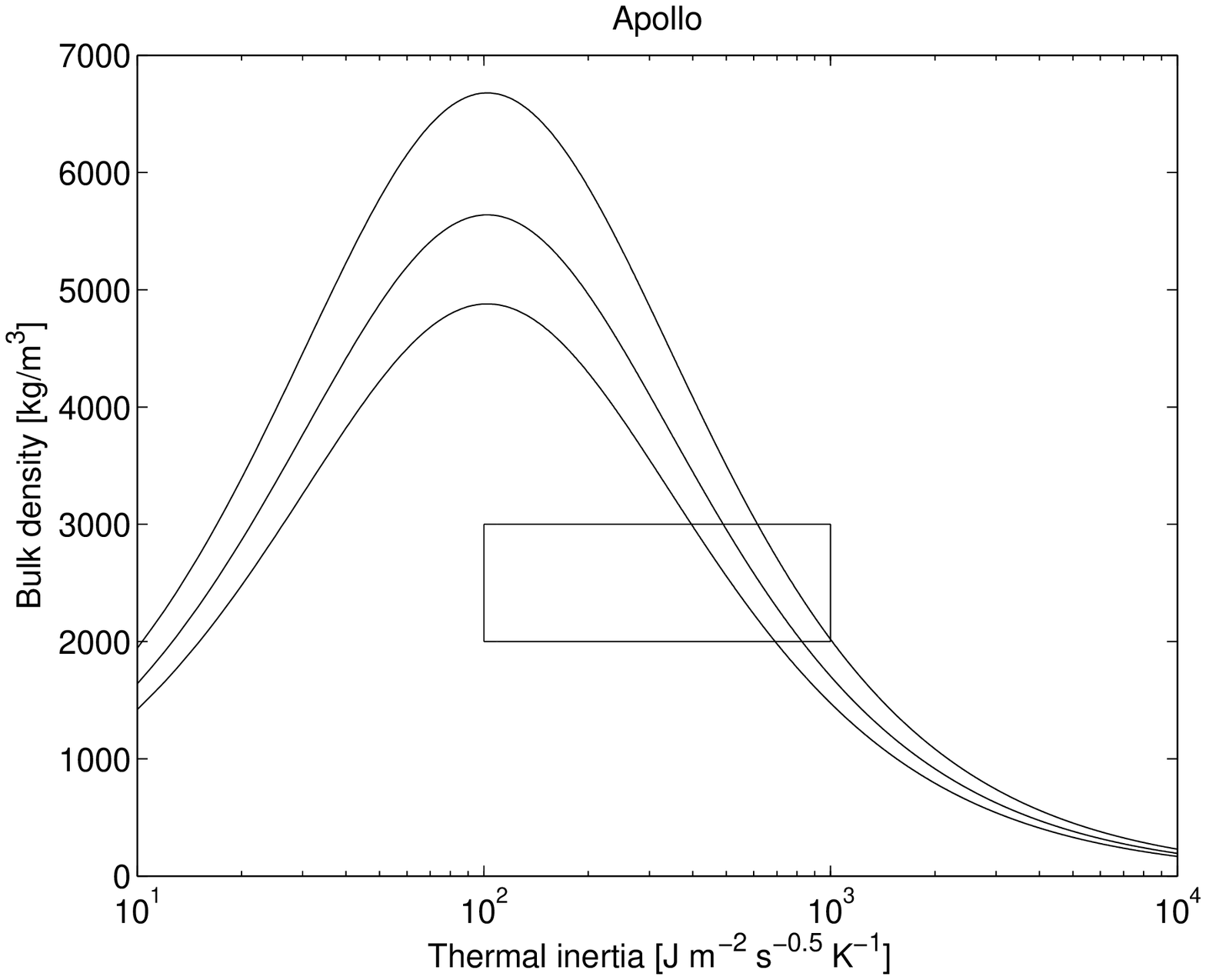}
   \includegraphics[width=6.8cm]{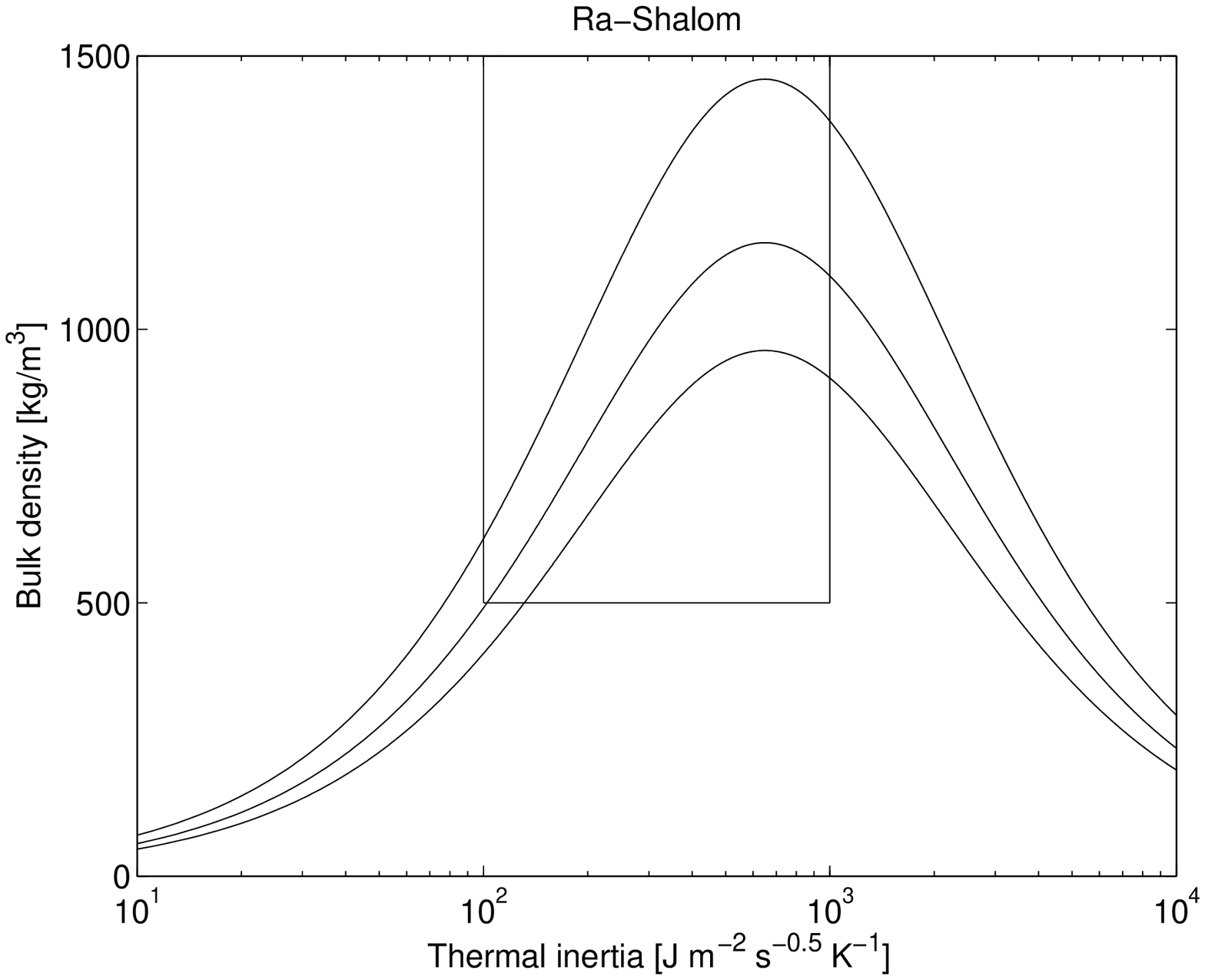}\includegraphics[width=6.8cm]{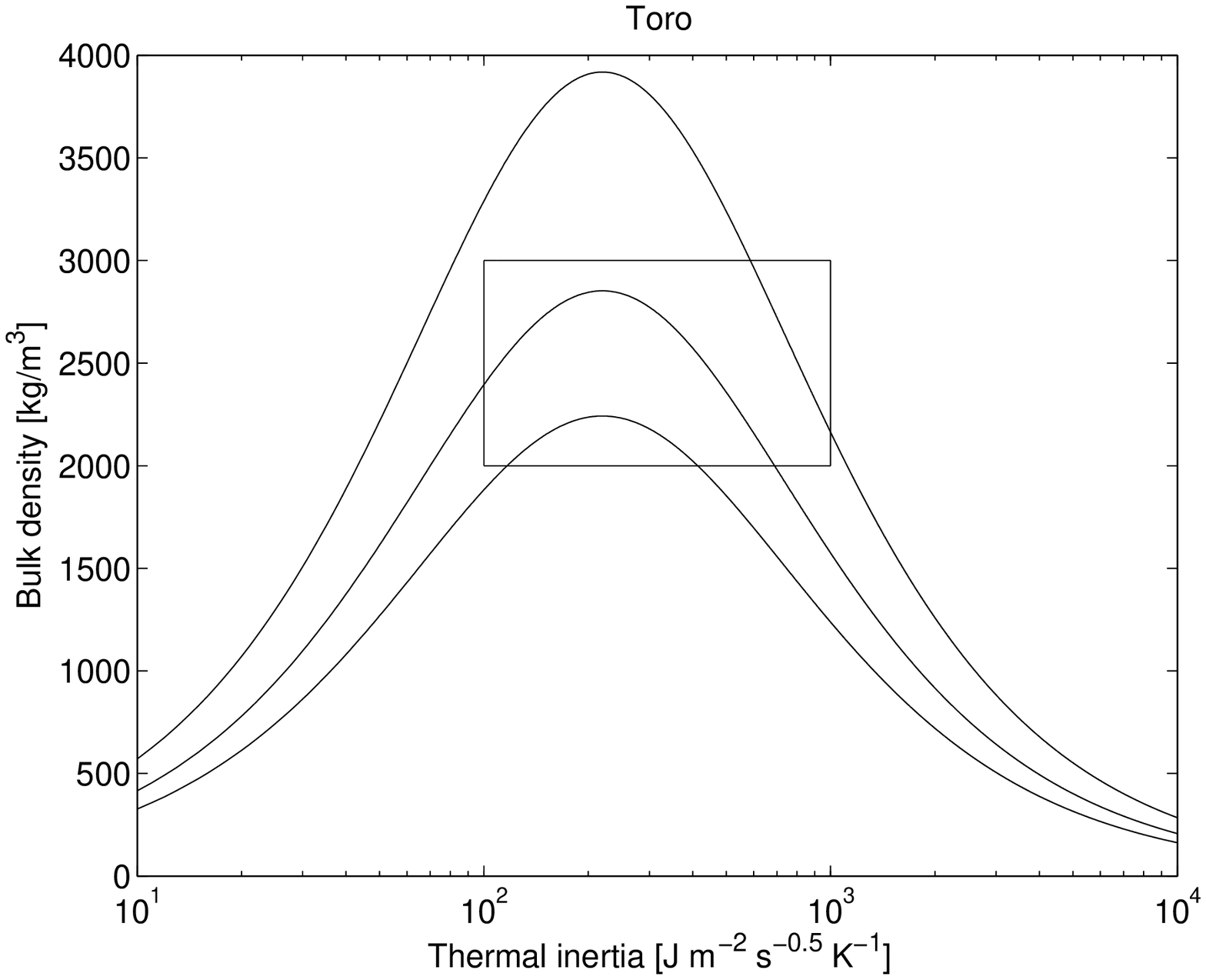}
    \includegraphics[width=6.8cm]{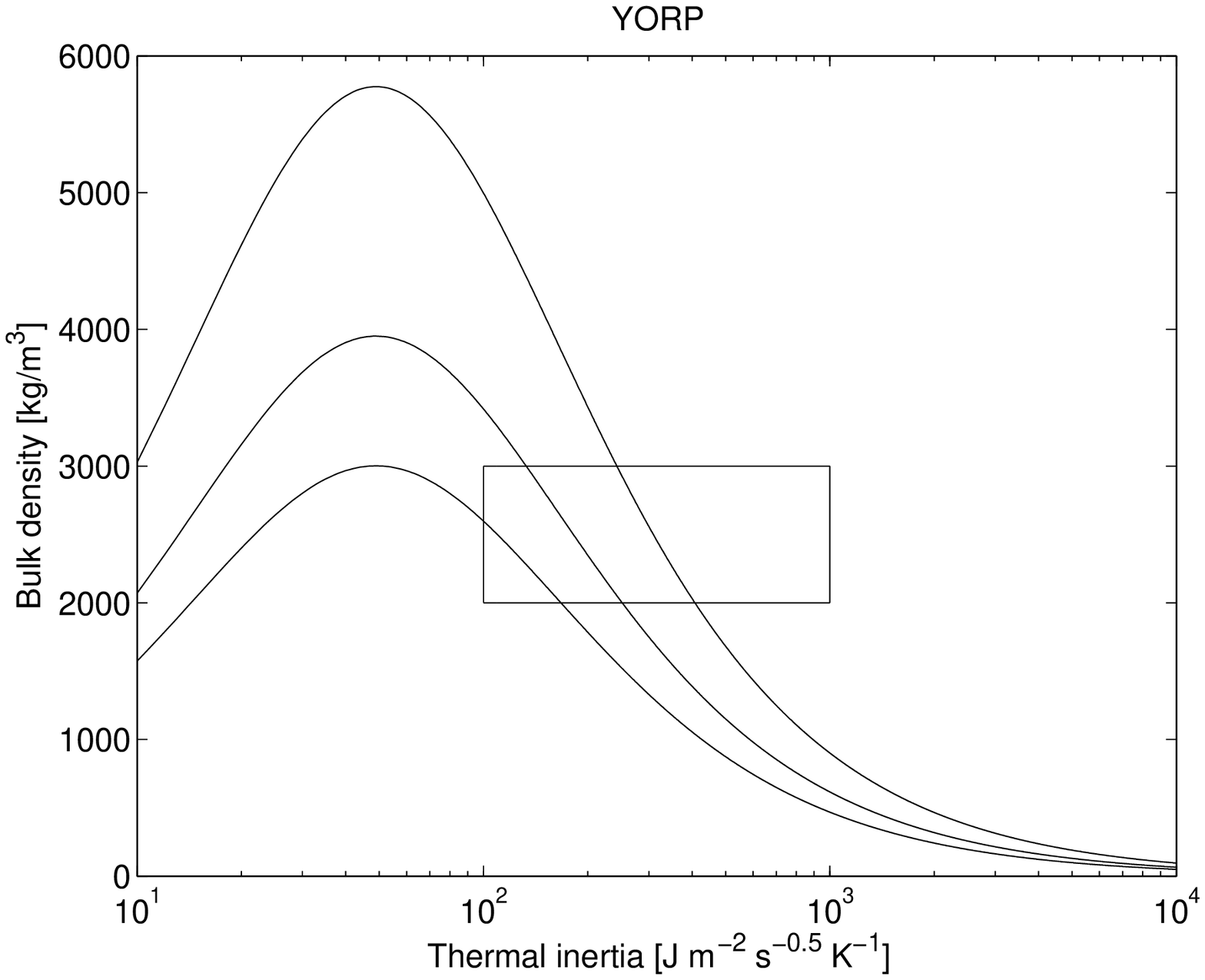}\includegraphics[width=6.8cm]{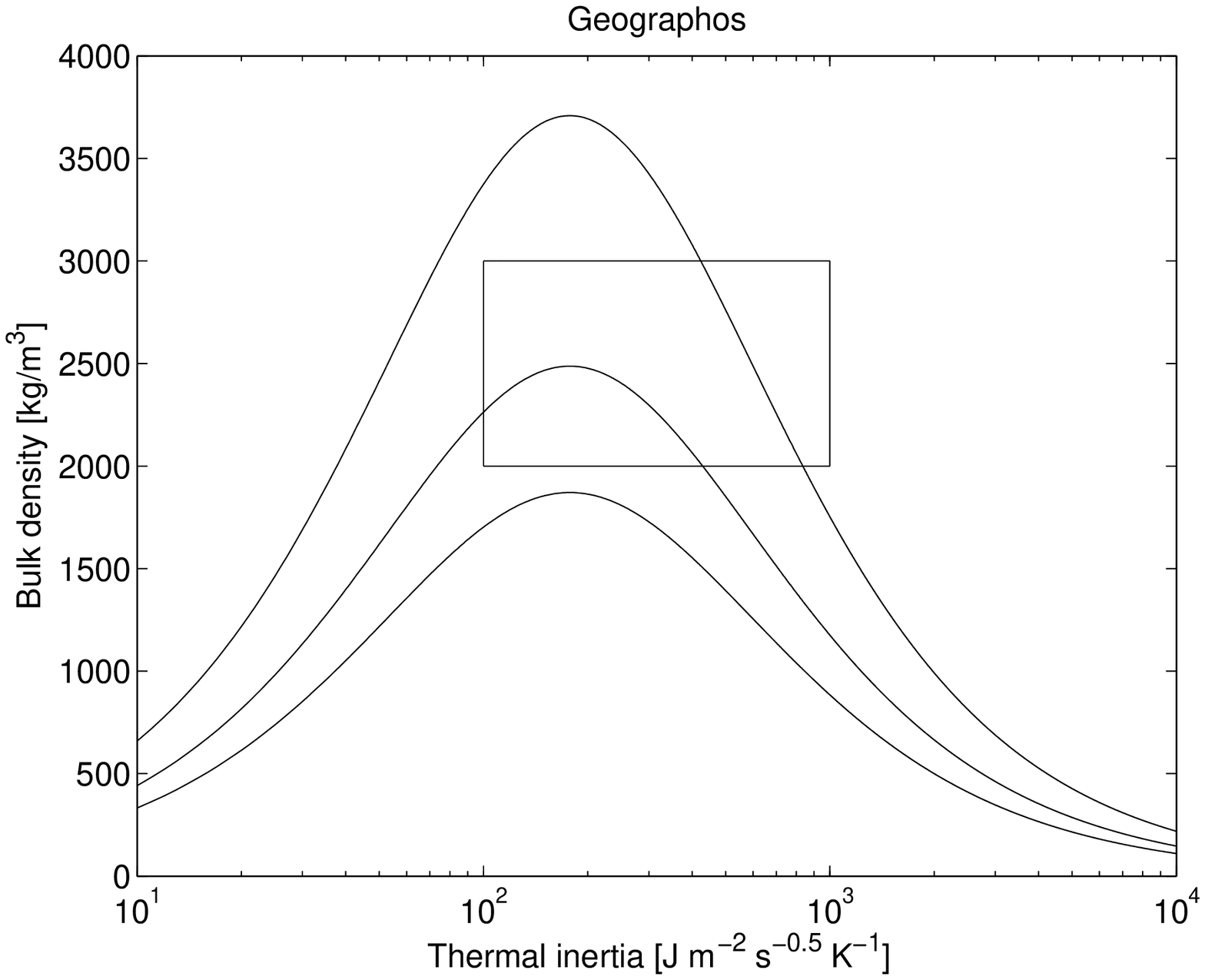}
    \caption{Density as a function of thermal inertia for asteroids
      Golevka, Apollo, Ra-Shalom, Toro, YORP, and Geographos. The
      rectangles correspond to reasonable values of $\rho$ according
      to the taxonomic type and to a reasonable range of $\Gamma$
      \citep{delbo}.}
  \label{fig:phys_const}
\end{figure*}

The results reported in Table~\ref{tab:yarko_tab} can be used to
constrain physical quantities. When $A$, $D$, and $\gamma$ are known
Eq.~\eqref{eq:A2} provides a simple relationship between $\rho$ and
$\Theta$. This relationship can be easily translated into a
relationship between $\rho$ and the thermal inertia $\Gamma$ by means
of Eq.~\eqref{eq:thermal}. As a benchmark of these technique we can
use asteroid 1999 RQ$_{36}$ (Fig.~\ref{fig:rq36}), for which the known
values of $\Gamma$ and $\rho$ \citep{RQ36} match the plotted
contraint.
Figure~\ref{fig:phys_const} shows the possible values of $\rho$ as a
function of $\Gamma$ for asteroids Golevka, Apollo, Ra-Shalom, Toro,
YORP, and Geographos. For the latter two objects we assumed slope
parameter $G=0.15$.  Our findings are consistent with the taxonomic
type. For instance, Golevka, Apollo, Toro, YORP, and Geographos are
S/Q-type asteroids with an expected density between 2000 and 3000
kg/m$^3$, while Ra-Shalom is a C-type so we expect a bulk density from
500 and 1500 kg/m$^3$. Figure~\ref{fig:phys_const} suggests that
Golevka has thermal inertia $150 <\Gamma < 500$ J m$^{-2}$ s$^{-0.5}$
K$^{-1}$ and Apollo has a rather large thermal inertia $400 <\Gamma <
1000$. According to \citet{delbo_ra-shalom} Ra-Shalom has an unusually
high thermal inertia. In fact, by taking the right side of the plotted
rectangle we obtain a density closer to 1000 kg/m$^3$, similar to the
one 1999 RQ$_{36}$, which belongs to a similar taxonomic class. For
Toro, \citet{mueller} reports a thermal inertia $200 < \Gamma < 1200$
J m$^{-2}$ s$^{-0.5}$ K$^{-1}$ but likely lower, which would result in
a bulk density between 2000 and 4000 kg/m$^3$.

\subsection{Implications for impact predictions}\label{s:impact}
The question is how much the predictions of possible asteroid impacts
on Earth can be affected by the Yarkovsky effect.  There are already
three well known examples, namely the asteroids Apophis, 1999
RQ$_{36}$, and 1950 DA for which Yarkovsky perturbations are relevant
and need to be accounted for in the impact risk assessment.

For Apophis and 1999 RQ$_{36}$ this is due to the presence of a
strongly scattering planetary close approach between now and the
epochs of the possible impacts. These encounters transform a very well
determined orbit into a poorly known one for which chaotic effects are
dominant \citep{milani_rq36}. Apophis will have a close approach in
April 2029 with minimum distance of $\sim$ 38000 km from the
geocenter. As a consequence the orbital uncertainty will increase by a
factor $>40000$. 1999 RQ$_{36}$ will have a close approach to Earth in
2135 with nominal minimum distance about the same as the distance to
the Moon, with an increase in uncertainty by a factor $\sim$500. The
minimum possible distance for this close approach is three times
smaller and would result in an increase of the uncertainty by a factor
$\sim$1500 \citep{RQ36}. In both cases the Yarkovsky effect is large
enough to shift the position at the scattering close approach by an
amount much larger than the distance between the keyholes
\citep{chodas} corresponding to impacts in later years (2036, 2037,
2068 for Apophis; 2175, 2180, 2196 for 1999 RQ$_{36}$). Thus, the
occurrence of these later impacts is determined by the Yarkovsky
perturbation in the years between now and the scattering encounter.
For 1950 DA the influence of the Yarkovsky effect for the possible
impact in 2880 is due to long time interval preceding the impact that
allows the orbital displacement accumulate \citep{giorgini02}.

Currently, 1999 RQ$_{36}$ is the case with the best determined
Yarkovsky effect (SNR $\sim$ 200), while Apophis and 1950 DA have only
a marginal detection (SNR $<$ 1 and SNR $\sim$ 1.4,
respectively). Therefore, the impact monitoring for 1999 RQ$_{36}$
fully takes into account Yarkovsky \citep{RQ36}. On the contrary, the
current estimate of impact probabilities for Apophis is based on a
Montecarlo model of Yarkovsky based on a priori knowledge of the
statistical properties of this effect \citep{apophis}. 1950 DA could
benefit from a similar approach. The knowledge of the Yarkovsky
perturbation for Apophis is expected to increase dramatically due to
observations possible during the January 2013 close approach.

We investigated the possibility that our identification of asteroids
with measurable Yarkovsky produces new cases such as the two above,
that is of impact monitoring affected by Yarkovsky. The answer to this
question is negative, in that the intersection between the current
list of NEA with possible impacts on Earth (337 according to NEODyS,
404 according to Sentry) and the list with detected Yarkovsky contains
only 1999 RQ$_{36}$.

However, this conclusion depends on the fact that our monitoring of
possible future impacts is done for only about one century (currently
90 years for NEODyS, 100 for Sentry). 1999 RQ$_{36}$ was a special
case, related to an intensified effort for the OSIRIS-REx mission
target 1999 RQ$_{36}$ (http://osiris-rex.lpl.arizona.edu). If this
time span were generally increased to 150--200 years, there could well
be other cases similar to 1999 RQ$_{36}$.

For asteroid 1950 DA \citet{busch} report a density around $3000$
km/m$^3$ and two possible solutions for
pole orientation and effective diameter: 
\begin{enumerate}
\item $\gamma = 24.47^\circ$ and $D = 1.16$ km;
\item $\gamma = 167.72^\circ$ and $D = 1.30$ km.
\end{enumerate}
By scaling from the 1999 RQ$_{36}$ case, we obtain $A_2 = 7.01$
au/d$^2$ for the direct solution and $A_2 = -5.83$ au/d$^2$ for the
retrograde solution. Even if we found a low SNR detection, our result
strongly favors the retrograde solution, which is at 0.6 $\sigma$,
than the direct solution, which is more than 3 $\sigma$ away. As a
consequence, the 2880 impact would be ruled out \citep{giorgini02}.

\section{Conclusions}
In this paper we developed a 1-parameter formulation that models the
transverse component of the Yarkovsky effect. Despite being simple,
this formulation captures the essence the Yarkovsky effect and also
allows us to solve for the introduced parameter as a part of the
orbital fit.

The small magnitude of the Yarkovsky acceleration made it necessary to
employ a high precision dynamical model including the effects of
sixteen perturbing asteroids and planetary relativistic terms.

As the solution of the orbital fit depends on the observational data
treatment, we applied the debiasing and weighting scheme by
\citet{cbm10}. Moreover, we resorted to some manual weighting in cases
where an incorrect treatment non-CCD observations could have led to
unreliable results.

After filtering out those objects with an unreasonably high orbital
drift, we obtained 21 robust detections with a signal to noise ratio
grater than 3. By connecting the sign of the orbital drift to the spin
orientation, we found a 4:1 fraction of retrograde rotators. We
discussed how this excess of retrograde rotators can be related to the
delivery of NEAs to the inner Solar System.

For a few asteroids we were able to use the measured orbital drift
along with the known physical characterization to constrain unknown
physical quantities such as bulk density and thermal inertia.

We also discussed the implication of the Yarkovsky effect and pointed
out how this perturbation can be relevant either when there is a
scattering close approach before the possible impact or when the
impact is far enough in time.

\section*{Acknowledgments}
DF was supported for this research in part by an appointment to the
NASA Postdoctoral Program at the Jet Propulsion Laboratory, California
Institute of Technology, administered by Oak Ridge Associated
Universities through a contract with NASA, and in part by
ESA/ESTEC-SpaceDyS Service Level Agreement
SSA-NEO-ESA-SLA2-002\_NEODYS.

SC conducted this research at the Jet Propulsion Laboratory,
California Institute of Technology, under a contract with the National
Aeronautics and Space Administration.

The work of DV was partially supported by the Czech Grant Agency
(grant 205/08/0064) and Research Program MSM0021620860 of the Czech
Ministry of Education.

\bibliography{yarko}
\bibliographystyle{elsarticle-harv}
\end{document}